\documentclass[aps, nofootinbib,superscriptaddress]{revtex4}
\usepackage{psfrag}
\usepackage{array}
\usepackage{amstext,amsmath,amssymb,amsfonts,bbm}
\usepackage[latin1]{inputenc}
\usepackage[dvips]{graphicx}
\usepackage{epsfig}

\newcommand{\N}{\mathbb{N}}

\newcommand{\C}{\mathbb{C}}

\newcommand{\f}{\frac}

\newcommand{\rar}{\rightarrow}

\newcommand{\SU}{\mathrm{SU}}

\def\vphi{\varphi}
\def\la{\langle}
\def\ra{\rangle}

\newcommand{\cG}{{\cal G}}
\newcommand{\cH}{{\cal H}}

\newcommand{\cN}{{\cal N}}
\newcommand{\cO}{{\cal O}}

\newcommand{\id}{\mathbb{I}}

\def\eps{\epsilon}

\def\mone{^{-1}}

\newcommand{\be}{\begin{equation}}
\newcommand{\ee}{\end{equation}}
\newcommand{\bes}{\begin{eqnarray}}
\newcommand{\ees}{\end{eqnarray}}

\newcommand{\Ref}[1]{(\ref{#1})}

\newcommand{\sixj}[2]{\left\{\begin{array}{ccc} #1 \\ #2 \end{array}\right\}}
\def\sj{$\{6j\}$}
\def\nj{$\{9j\}$}

\def\vtheta{\vartheta}

\def\pp{\partial}
\usepackage{color}

\begin{document}

\title{\large\bf Recurrence relations for spin foam vertices}

\author{Valentin Bonzom}\email{valentin.bonzom@ens-lyon.fr}
\affiliation{Laboratoire de Physique, ENS Lyon, CNRS-UMR 5672, 46 All\'ee d'Italie, Lyon 69007, France}
\affiliation{Centre de Physique Th\'eorique, CNRS-UMR 6207, Case 907 Luminy, Marseille 13288, France}

\author{Etera R. Livine}\email{etera.livine@ens-lyon.fr}
\affiliation{Laboratoire de Physique, ENS Lyon, CNRS-UMR 5672, 46 All\'ee d'Italie, Lyon 69007, France}

\author{Simone Speziale}\email{speziale@cpt.univ-mrs.fr}
\affiliation{Centre de Physique Th\'eorique, CNRS-UMR 6207, Case 907 Luminy, Marseille 13288, France}

\date{\small \today}

\begin{abstract}
We study recurrence relations for various Wigner 3nj-symbols and the non-topological 10j-symbol. For the 6j-symbol and the 15j-symbols which correspond to basic amplitudes of 3d and 4d topological spin foam models, recurrence relations are obtained from the invariance under Pachner moves and can be interpreted as quantizations of the constraints of the underlying classical field theories. We also derive recurrences from the action of holonomy operators on spin network functionals, making a more precise link between the topological Pachner moves and the classical constraints.
Interestingly, our recurrence relations apply to any $\SU(2)$ invariant symbol, depending on the cycles of the corresponding spin network graph. Another method is used for non-topological objects such as the 10j-symbol and pseudo-isoceles 6j-symbols. The recurrence relations are also interpreted in terms of elementary geometric properties. Finally, we discuss the extension of the recurrences to take into account boundary states which leads to equations similar to Ward identities for correlation functions in the Barrett-Crane model.


\end{abstract}

\maketitle

\section{Introduction: Spinfoams and the BC vertex}
\label{intro}

Spin foam amplitudes $\cal A$ provide a non-perturbative and background independent definition of the path integral for general relativity.
Considering a four-dimensional manifold with a triangulated boundary $\Sigma$ provided with a given classical (discrete) metric $q_{ab}$ (on $\Sigma$), then the spin foam proposal is:
\be\label{A=K}
{\cal A}[q_{ab}] \sim K[q_{ab}] \equiv \int_{q_{ab}}{\cal D}g_{\mu\nu} \, e^{i S_{\rm GR}[g_{\mu\nu}]}.
\ee
In particular, a key property of the right hand side is to project on the kernel of the spatial diffeomorphisms ${\cal H}_a$ and Hamiltonian constraint $\cal H$ present in the action, i.e. to satisfy the set of formal equations
\be\label{WdW}
\hat{\cal H}_a(q_{ab}, \delta/\delta q_{ab}) \, K[q_{ab}] = 0, \qquad
\hat{\cal H}(q_{ab}, \delta/\delta q_{ab}) \, K[q_{ab}] = 0.
\ee
These Wheeler-DeWitt equations encode at the quantum level the invariance under space-time diffeomorphisms of general relativity.
%
%
Therefore one of the requirements of a consistent spin foam model is to implement some version of these equations.
In its present formulation, a spin foam model is defined as a sum over bulk triangulations,
\be\label{Asf}
{\cal A}[q_{ab}] = \sum_v \lambda^v c_v[q_{ab}]
\ee
where $v$ is the number of 4-simplices of the triangulation and $\lambda$ a dimensionless coupling constant -- the notation $v$ comes from the fact that 4-simplices are dual to vertices in the 2-complex dual to the triangulation. This justifies the name ``vertex expansion'' for the sum \Ref{Asf}.
The amplitudes $c_v$ here include a sum over the triangulations with the same $v$ and are built with an elementary amplitude assigned to each 4-simplex. The amplitudes are algebraic quantities from the representation theory of the local Lie group of general relativity. A key result of this construction is to represent the boundary metric in terms of \emph{discrete} quantities, i.e. piecewise flat metrics with discrete values, typically half-integers. This provides a match with the labels of spin network states, the complete basis of the kinematical Hilbert space of LQG. Spin foam models can then be used to give transition amplitudes to spin networks. To support the conjecture that the amplitudes represent quantum gravity, one can study equations like \Ref{WdW} and in particular link the operator $\hat {\cal H}$ to some version of the Hamiltonian constraint of general relativity. A priori, the connection with LQG suggests that $\hat {\cal H}$ should be an operational version of Ashtekar's Hamiltonian ${\cal H} \sim EEF$, however such a link is still missing for existing spin foam models (see however \cite{AlesciNoui,Zapata,BiancaJ,Bianca,BiancaB}).

On the other hand, there are semiclassical approximations of spin foam models which suggest a different direction to interpret $\cH$. In these approximations, the piecewise flat metrics $q_{ab}$ are seen as Regge geometries, and the amplitudes $c_v[q_{ab}]$ are given by exponentials of the Regge action, thus providing a clear link with general relativity. In this approximation, \Ref{WdW} should be realized with $\hat\cH$ being an operatorial version of the Hamiltonian for Regge calculus. Unfortunately the construction of the canonical framework for Regge calculus is still under development \cite{Zapata,Bianca,BiancaJ,BiancaB}, and there is no clear proposal for the Hamiltonian.
The difficulty of representing the Hamiltonian constraint in these variables has two origins. The first one comes from the discretization of the boundary manifold. Schematically, the form of the classical constraint is
\be\label{Hcl}
\cH \sim K^2 + R,
\ee
where $K$ and $R$ are respectively the extrinsic and intrinsic curvature of the 3d manifold. In a classical simplicial context, these two quantities are discretized respectively on $n-1$ and $n-2$ simplices. Therefore different realization of $\cH$ can be a priori envisaged.
The second difficulty comes from the fact that on top of being a Regge geometry as opposed to a continuum one, $q_{ab}$ in \Ref{Asf} is also represented by discrete quantities such as half-integers. Hence the version of \Ref{WdW} we are after should be a difference equation. This second difficulty offers a new light on investigating the issue. In fact, difference equations arise naturally in the representation theory of Lie groups used to build the amplitudes in \Ref{Asf}, thus it is tempting to look there for an implementation of the idea of representing the Wheeler-de Witt equations as difference equations.

There are two instances where this idea is at least partially realized. The first one is loop quantum cosmology (see \cite{lqc} and references therein). There the mini-superspace setting allows a construction of the Ashtekar Hamiltonian $\cH\sim EEF$ (which unambiguously lives on $n-2$ simplices) precisely as a difference equation. In turn, the exponentiated action has been recently argued \cite{AbhaySF} to give rise to a vertex expansion of the form \Ref{Asf}.

The second instance is the Ponzano-Regge model for 3d Euclidean quantum gravity. Recurrence relations for the 6j-symbol are well-done \cite{schulten1,schulten2}. They can be used to derive the asymptotics of the 3d quantum gravity amplitudes at large scales \cite{schulten2,maite} and for fast numerical computations of the 6j-symbol.
We begin our paper with a discussion of recurrence relations for this model and their connection to Wheeler-de Witt equations, Section \ref{section:6j}. We discuss various techniques to derive recurrence relations, which allow us to reobtain the existing ones and to produce new ones. One of these methods is very standard, making use of the topological invariance of the model. A second method gives a clear relation between the classical constraints and the 1-4 Pachner move, using the action of holonomy operators on spin network functionals (as a simple case of the analysis of \cite{ooguri3d}). Finally, for the special case of isosceles 6j-symbols, we present a new technique which will be generalized to the 10j-symbol of the non-topological Barrett-Crane model.

Then we tackle the issue of four dimensional models. In Section \ref{fifj-recurrences}, we consider Ooguri's model \cite{ooguri}. This is a topological model corresponding to the quantization of BF theory and not to quantum gravity, but is often used as a starting point to build spin foam models for 4d gravity (see \cite{epr} for the construction of the most recent spin foam models). We show that the topological invariance again encodes difference equations for the $\SU(2)$ 15j-symbols. Like in 3d, such relations provide a connection between the classical flatness constraint $F=0$ and the Pachner moves.

For the above topological models, this shows that the invariance under Pachner moves contains the dynamics generated by the underlying classical symmetries. These symmetries are such that for a given topology, all bulk triangulations contribute to the sum \Ref{Asf} with the same amplitude (upon regularization). Thus, the projection onto the kernel of the constraints is still realized when the sum \Ref{Asf} is restricted to a single triangulation, say to the first order in $\lambda$. However, this is not the case anymore for non-topological models. In such models, we expect the classical symmetries to be restored, and the spin foam amplitude $\mathcal{A}[q_{ab}]$ to project on the kernel of the constraints only through the sum over the bulk triangulations for a given boundary. But we do not know at the present day how to get closed expressions for non-trivial triangulations.

Nevertheless, this is not the end of the story, and it is certainly interesting to look at difference equations even for the first order  in $\lambda$. If we formally think of the projector $\mathcal{A}$ as the exponentiation of the Diffeomorphism and Hamiltonian constraints, and of the sum over spin foams as a sum over histories of spin networks, then we expect the first order of \Ref{Asf} to be related to the matrix elements of the constraints. As already mentioned, there are difficulties in representing the constraints. Thus, recurrence relations are certainly useful to probe their basic properties. In this regard, we emphasize in this article the geometric interpretation of the recurrence relations we obtain. In particular, for the topological models mentioned  above, we found that the basic recurrences are generated by gluing a flattened simplex to an initial one. From the point of view of the latter, it results in an elementary displacement of a point, which induces shifts of lengths or areas.

Recurrence relations could also put restrictions on the form of the admissible higher orders and help the renormalization process. Indeed, if a correction of order $n$ violates a difference equation, it implies that it does not fulfill some geometric property of the initial amplitude, and thus might be irrelevant for renormalization.

With these perspectives in mind, we finally focus on the Barrett-Crane model \cite{BC1} in Section \ref{BCrec}. This is a non-topological model which assigns an amplitude, called the 10j-symbol, to any 4-simplex whose areas take discrete values. Although it has been shown not to provide a correct quantization of general relativity, it captures some important features provided we only look at a single simplex. We derive a difference equation of order 4 for the 10j-symbol, which is solved in the large area limit by the cosine of the Regge action. This leads to a nice geometric interpretation showing that the recurrence probes the closure of the 4-simplex. Then, we propose to look at difference equations not for $\mathcal{A}$ alone, but together with a boundary state. Such a state should be thought of as a physical state, satisfying the Wheeler-DeWitt equations for discrete metrics. By peaking such a state on a well-defined boundary geometry, we may expect to suppress the non-geometric parts of the recurrence relations, which are for instance associated to the degenerate configurations of the 10j-symbol \cite{asympt,asymptlaurent}. This is precisely the strategy used in semi-classical computations of observable correlations in the Barrett-Crane model \cite{simone1}. In this context, the recurrence relations moreover translate into some Ward-like identities for insertions of observables, as was recently explained in the 3d case in \cite{maite}.

\section{Recurrence relations for $\{6j\}$-symbols} \label{section:6j}
It is well-known that the 6j-symbol satisfies the following recurrence relation,
\be\label{rec6j}
A_{+1}(j)\,\begin{Bmatrix} l_1 &l_2 &l_3 \\ j_1+1 &j_2 &j_3 \end{Bmatrix} +
A_{0}(j)\,\begin{Bmatrix} l_1 &l_2 &l_3 \\ j_1 &j_2 &j_3 \end{Bmatrix} +
A_{-1}(j)\,\begin{Bmatrix} l_1 &l_2 &l_3 \\ j_1-1 &j_2 &j_3 \end{Bmatrix} = 0.
\ee
The exact values of the coefficients are not important to our considerations here, but the interested reader can found them in the Appendix. This recurrence relation is a second order difference equation for one of the spins (here chosen to be $j_1$, but any other can be chosen suitably changing the coefficients).
This is one of many recurrence relations satisfied by the 6j-symbol (see e.g. \cite{varsha} for examples), which can all be derived from the Biedenharn-Elliott identity. The latter (together with the orthogonality relation and a basic associativity property, see \cite{schulten1}) fixes uniquely the properties of the 6j-symbol, hence it can be seen as its defining relation.
In particular, it can be used to prove that a state sum model for three dimensional quantum gravity based on a tetrahedral amplitude given by the 6j-symbol is a topological invariant.
The Biedenharn-Elliott identity turns out to be precisely the statement of the invariance under the so-called 2-3 Pachner move for that model.
It thus intimately related to the special homeomorphism symmetry of three dimensional gravity. Since this is in turn the symmetry imposed by the classical constraints, one can argue that the recurrence relations capture a quantum version of the Wheeler-de Witt equation \Ref{WdW}.

A connection between \Ref{rec6j} and a partial differential equation can be made taking the continuum limit. Let us define $f(j_l) = \sqrt{12\pi V(\ell_l)}\,\{6j\} $, where $V$ is the volume of the tetrahedron $\tau$ whose edge lengths are given by $\ell_l=j_l+1/2$, and corresponding dihedral angles by $\theta_l(j_l)$.
If we recale homogeneously the spins, i.e. $j_i\mapsto Nj_i$, and send $N\rightarrow \infty$, the relation \Ref{rec6j} is turned into the following second order difference equation \cite{schulten2}:
\be\label{Hlink}
\left[\Delta_{j_1} + 2-2\cos\theta_1(j_l) \right]\ \f 1{\sqrt{\sin\theta_1(j_l)}}f(j_l)= 0,
\ee
where the difference operator $\Delta$ is defined through: $\Delta f(x) = f(x+1)+f(x-1)-2f(x)$, and $\theta_1$ is the dihedral angle between the two triangles meeting at the edge carrying the spin $j_1$ in $\tau$, computed from the edge lengths $\ell_l$. In the large spin limit this difference equation is approximated by a second order partial differential equation whose solution
leads to the known asymptotics of the 6j-symbol in terms of the Regge action\footnote{In general, the solution is combination of Airy functions, which reduces to \Ref{6jasymp} when $V(\ell_e)^2>0$.} $S_{\rm R}$ \cite{Ponzano,schulten2},
\be\label{6jasymp}
\begin{Bmatrix} j_1 &j_2 &j_3 \\ j_4 &j_5 &j_6 \end{Bmatrix} \sim
\f1{\sqrt{12\pi V(\ell_l)}}\cos\left( S_{\rm R}[\ell_l] + \f\pi4\right).
\ee
The large spin regime then provides a notion of semiclassical limit in which the spin foam amplitudes are approximated by quantum Regge calculus, see \cite{graviton3d,valentin}.

The asymptotic expression \Ref{Hlink} of the recurrence relation suggests to look for a discretization of the classical Hamiltonian constraints on the links of the triangulation, as $p_l = \theta_l(j_l)$ \cite{Bianca,BiancaJ}. This can be seen as the realization of all the constraints (not just the Hamiltonian \Ref{Hcl}) as $F=0$, a form peculiar to 3d gravity.

However, apart from this allusive analogy, a more explicit connection between \Ref{rec6j} and a discrete version of the WdW equation has never been achieved so far (see also \cite{BarrettWdW}).
The difficulty lies in the difficulties with a canonical version of regge calculus, and on the question of whether the constraints should be discretized.

A simplification can be obtained studying homogeneous configurations, i.e. taking all the spins equal to $j$ and studying what happens under variations of $j$. No analytic recurrence relation can be built in terms of $j$ alone (we will come back to this point below). Nevertheless in the large spin limit we can use the explicit solution \Ref{6jasymp} to derive the following equation,
\be\label{omo1}
\left[ \pp^2_j+ (6\theta_0)^2 \right] f(j) = 0,
\ee
where $\theta_0=\arccos(1/3)$ is the dihedral angle of an equilateral tetrahedron. A natural interpretation for this equation is a cosmological model. The amplitude associated to the tetrahedron can be seen a the Hartle-Hawking state for a two-sphere ${\cal S}^2$ discretized by four triangles. In fact any higher triangulation in the bulk would be divergent because of a gauge symmetry, and fixing the corresponding gauge degrees of freedom following \cite{PR1} would reduce the bulk triangulation to just the trivial one of a single tetrahedron.
This is of course a model which is way too simple to be meaningful,\footnote{The model can be made less trivial considering the Turaev-Viro model, because a non-vanishing cosmological constant introduces an extra quadratic dependence in the Wheeler-DeWitt equation. However we will not pursue this line further.
}
and one should at least extend these considerations to a more complicated boundary triangulation and the resulting spin foam amplitude. A systematic analysis of recurrence relations could be useful to this end.

We now present a new type of recurrence relations, that can be obtained using an integral representation of the 6j -symbol. The latter can be given for a special ``isosceles'' configuration of the 6j-symbol. For a general configuration one needs to take the square of the symbol. We will not deal with this case.

We now consider pseudo-isoceles 6j-symbols, which admit an integral formulation. It makes it easy to derive a new recurrence relation. We study its relation to the Biedenharn-Elliott identity, and emphasize the underlying geometric content.

\subsection{Recurrence relations for the Isoceles $\{6j\}$-symbol}

The so-called ``isoceles" 6j-symbols, relevant to the calculation of graviton-like correlation in 3d
quantum gravity \cite{graviton3d,valentin}, admits a simple integral formulation,
\be\label{6jIsoInt}
F(j,k,J,K)\,\equiv\, \int dg dh \chi_j(g)\chi_k(h)\chi_J(gh)\chi_K(gh^{-1})
\,=\, (-1)^{2j} \sixj{j &J &K}{k &J &K},
\ee
where $\chi_j(g)=\sin d_j\theta/\sin\theta$ is the character of the $\SU(2)$ representation of spin $j$ evaluated at the
group element $g$ with class angle $\theta$. Here $d_j=(2j+1)$ is the dimension of the representation.
Notice that $(-)^{2j} = (-)^{2k}$ because of the parity condition on the two triplets of
representations $(j+J+K)\in\N$ and $(k+J+K)\in\N$.

Thanks to the symmetries of the integrand, the two integrals over SU(2) can be reduced to a single triple integral. This is most conveniently written in terms of four angles subjected to a constraint. Let us call $\alpha$ and $\beta$ the class angles of $g$ and $h$, and $\phi^\pm$ those of the products $gh^{\pm}$. Then we can write \Ref{6jIsoInt} as an integral over these four angles with the constraint
\be\label{6jIsoC}
\cos\alpha\,\cos\beta\,=\,\f12(\cos\phi^++\cos\phi^-).
\ee
Observe that \Ref{6jIsoC} can be written in terms of the characters of the fundamental representation as
\be
C(g,h) = \chi_{\f12}(g)\chi_{\f12}(h) - \chi_{\f12}(gh)+\chi_{\f12}(gh^{-1}) = 0.
\ee
Inserting $C(g,h) $ in the integral \Ref{6jIsoInt} gives automatically zero. On the other hand, using the recoupling formula
\be
\chi_j(g)\,\chi_{\f12}(g) \,=\, \chi_{j-\f12}(g)+\chi_{j+\f12}(g)
\ee
we can split the integral into a sum of different isosceles 6j-symbols. This method allows us to immediately
derive the following recurrence relation,
\be \label{recurrenceiso6j}
0 = \sum_{\eps=\pm}\sixj{j &J+\f\eps2 &K}{k &J +\f\eps2 &K}+\sum_{\eps'}\sixj{j &J &K+\f{\eps'}2}{k &J
&K+\f{\eps'}2} +
\sum_{\eta,\eta'}\sixj{j+\f\eta2 &J &K}{k+\f{\eta'}2 &J  &K}.
\ee

We would like to point out the simplicity with which we derived this relation. As mentioned above, all properties of the 6j-symbol are encoded in the Biedenharn-Elliott identity, thus also this one can be derived from it, as we show below. However it is remarkable to notice that having an integral representation at disposal paves the way to natural and straighforward derivations of recurrence relations.

Unlike the standard recurrence relations found in the literature, this one has trivial coefficients and all the spins are shifted. These features are achieved at the price of increasing the number of the terms in the relation, from the usual three or four to eight. As a difference equation it is second order, since we are dealing with the sum and not the difference of terms involving variations of the same variable.
Finally, notice that with the help of an additional iteration it can be put in a form involving spin shifts of 1 instead of $1/2$. Such spin shifts by integer step might be useful in some contexts, in particular they respect the parity constraints on the representations.

In the perspective discussed above, it would be interesting to study whether this equation can be compared with a discretization of the Wheeler-De Witt equation on the whole tetrahedron, and we leave this issue open for further work. Let us nonetheless point out that even if we
take all spins equal to $j$, \Ref{recurrenceiso6j} does not give rise to a second order equation in $j$.
Therefore also taking \Ref{recurrenceiso6j} as a starting point, a homogeneous equation like \Ref{omo1} can only be obtained approximately.


\subsection{Relation to the Biedenharn-Elliott identity}
As we pointed out above, all properties of the symbol are captured by the Biedenharn-Elliott identity, including the standard recurrence relations such as \Ref{rec6j}. Therefore, also the new \Ref{recurrenceiso6j} should be derived from it. This is indeed the case, as we now show. The derivation is a bit lengthy, but it endows \Ref{recurrenceiso6j} with a simple geometric meaning.


The Biedenharn-Elliott identity reads:
\be \label{be}
\sixj{j &h &g}{k &a &b}\,\sixj{j &h &g}{f &d &c} = \sum_l (-1)^{S+l}(2l+1)\,\sixj{k &f &l}{d &a &g}\,\sixj{a &d &l}{c &b &j}\,\sixj{b &c &l}{f &k &h}
\ee
where $S$ is the sum of the nine fixed spins. Geometrically, the 6j-symbol $\begin{Bmatrix}j &h &g \\ k &a &b \end{Bmatrix}$ can be seen as a tetrahedron whose edges are labelled by the representations, with lengths $l_j=j+\f{1}{2}$, and its triangles are $(j,h,g),\ (j,a,b),\ (g,k,a)$ and $(k,h,b)$. The property \eqref{be} thus says that the amplitude associated with three tetrahedra sharing the edge $l$ equals that of two tetrahedra glued along the triangle $(g,h,j)$: this exactly states the invariance of the Ponzano-Regge model under the 2-3 Pachner move.

\begin{figure}
\begin{center}
\includegraphics[width=8.5cm]{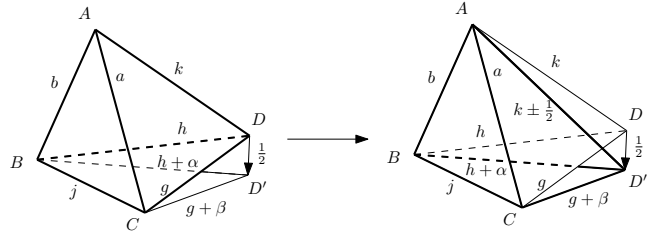}
\end{center}
\caption{The identity \Ref{recurrencebe}. On the left hand side, the tetrahedron $ABCD$ with edges $(g,h,j,k,a,b)$ is glued via the triangle $(g,h,j)$ to the flattened tetrahedron $BCDD'$ with an edge spin being $1/2$. On the right hand side, the three tetrahedra $ACDD'$, $ABDD'$ and $ABCD'$ share the edge $AD'$ with spin $k\pm\f12$. The tetrahedra $ACDD'$ and $ABDD'$ are ``flattened'' because of the spin $\f 12$ along $DD'$. Thus, from the point of view of the initial tetrahedron $ABCD$, the move can be seen as a ``small'' displacement of the point $D$ to $D'$, resulting in some small length shifts.
 \label{move2-3} }
\end{figure}

To recover the recurrence relation \Ref{recurrenceiso6j} from this identity, we first specialize $f=1/2$.
As a consequence, $d=g+\beta$ and $c=h+\alpha$ for $\alpha,\beta=\pm \f{1}{2}$, and the sum over $l$ reduces to two terms, $l=k\pm\f{1}{2}$:
\be \label{recurrencebe}
\sixj{j &h &g}{k &a &b}\,\sixj{j &h &g}{\f{1}{2} &g+\beta &h+\alpha} = \sum_{l=k\pm\f{1}{2}} (-1)^{S+l}(2l+1)\,\sixj{j &h+\alpha &g+\beta}{l &a &b}\,\sixj{a &g &k}{\f{1}{2} &l &g+\beta}\,\sixj{b &h &k}{\f{1}{2} &l &h+\alpha}.
\ee
The geometrical version of this relation is illustrated in Fig. \ref{move2-3}. The starting point of the move corresponds to adding a flattened tetrahedron to an initial one, so that the overall result can be seen as a ``displacement'' of the point $D$ to $D'$.

Consider the first two 6j-symbols on the right hand side: they correspond to the the tetrahedra $ABCD'$ and $ACDD'$, glued by the triangle $ACD'$ with edge spins $(a, g+\beta, l=k\pm1/2)$. We then apply the 2-3 move again, under the form \Ref{recurrencebe}, but now on the new tetrahedron $ABCD'$. Geometrically, a flattened tetrahedron $ACC'D'$ is glued along the triangle $ACD'$, and the 2-3 move transforms $ABCD'$ into $ABC'D'$, inducing shifts for the links meeting at $C'$.
We choose the shifts of the links $AC'$ and $C'D'$ to be $\alpha'=\alpha$ and $\beta'=-\beta$. The effect of this choice is to eliminate the shift of the spin $g$ due to the first move, while $a$ is shifted by $\alpha$, exactly like the opposite link $h$. Finally, each of the two terms of the r.h.s. of \eqref{recurrencebe} gives two terms, $BC$ becoming $BC'$ with the two possibilities $j\pm\f{1}{2}$.

These two successive moves lead to a relation between the five symbols:
\be
\sixj{j &h &g}{k &a &b},\ \quad\text{and}\quad \sixj{j+\f{\eta_j}{2} &h+\alpha &g}{k+\f{\eta_k}{2} &a+\alpha &b},\quad \text{for}\ \ \eta_j,\eta_k=\pm\f{1}{2}
\ee
and for a fixed $\alpha$. The coefficients of these relations are generically complicated. However, they become much simpler for the pseudo-isoceles case considered in the previous section. Take $h=a$ and $g=b$. Also redefine $a\rar a-\alpha$. The relations become for $\alpha=\pm\f{1}{2}$:
\begin{multline} \label{alpha1/2}
d_j d_k \sixj{j &a-\f{1}{2} & b}{k &a-\f{1}{2} & b} = \bigl(k+b-a+\f{1}{2}\bigr)\bigl(a+b-j+\f{1}{2}\bigr) \sixj{j-\f{1}{2} &a &b}{k-\f{1}{2} &a &b} - \bigl(k+b-a+\f{1}{2}\bigr)\bigl(a+b+j+\f{3}{2}\bigr) \sixj{j+\f{1}{2} &a &b}{k-\f{1}{2} &a &b} \\
+ \bigl(k+a-b+\f{1}{2}\bigr)\bigl(a+b-j+\f{1}{2}\bigr) \sixj{j-\f{1}{2} &a &b}{k+\f{1}{2} &a &b} - \bigl(k+a-b+\f{1}{2}\bigr)\bigl(a+b+j+\f{3}{2}\bigr) \sixj{j+\f{1}{2} &a &b}{k+\f{1}{2} &a &b}
\end{multline}
and:
\begin{multline} \label{alphamoins1/2}
d_j d_k \sixj{j &a+\f{1}{2} & b}{k &a+\f{1}{2} & b} = -\bigl(a+b+k+\f{3}{2}\bigr)\bigl(j+a-b+\f{1}{2}\bigr) \sixj{j-\f{1}{2} &a &b}{k-\f{1}{2} &a &b} - \bigl(a+b+k+\f{3}{2}\bigr)\bigl(j+b-a+\f{1}{2}\bigr) \sixj{j+\f{1}{2} &a &b}{k-\f{1}{2} &a &b} \\
+ \bigl(a+b-k+\f{1}{2}\bigr)\bigl(j+a-b+\f{1}{2}\bigr) \sixj{j-\f{1}{2} &a &b}{k+\f{1}{2} &a &b} + \bigl(a+b-k+\f{1}{2}\bigr)\bigl(j+b-a+\f{1}{2}\bigr) \sixj{j+\f{1}{2} &a &b}{k+\f{1}{2} &a &b}
\end{multline}

By symmetry, we get two similar relations with a shifted $b$ instead of $a$. Now, the sum of these four relations precisely gives the recurrence relation \eqref{recurrenceiso6j}. Notice first that the left hand sides of the above relations form exactly half of the relation, while the right hand sides involve the expected other terms. It then turns out that the sum, symmetrized in the exchange $(j\leftrightarrow k)$, indeed leads to trivial coefficients.

We have finally the following picture. Remember that the Biedenharn-Elliott identity is the key to the topological invariance of the Ponzano-Regge model. Equivalently, we can say that it generates and asks for the invariance of the vertex amplitude under the 2-3 Pachner move, i.e. under both the diffeomorphism and Hamiltonian constraints of Riemannian 3d gravity, since the Ponzano-Regge spin foam model implements these constraints on spin network states. Equations \eqref{alpha1/2} and \eqref{alphamoins1/2} thus generate and ask for the invariance under two successive moves, which deform the original tetrahedron by gluing some flattened tetrahedra. The relation \eqref{recurrenceiso6j}, which is weaker since it results from a sum of \eqref{alpha1/2} and \eqref{alphamoins1/2}, then implements the invariance under a combination of these moves, allowing for compensation between them.


\subsection{Recurrences from Holonomy Operators}\label{secRecHol}

\begin{figure}
\begin{center}
\includegraphics[width=7cm]{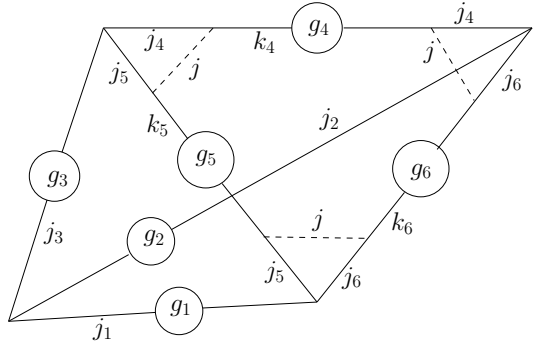}
\end{center}
\caption{The functional $\vphi_{\{j_l\}}(g_1..g_6)$ can be represented by a tetrahedron whose links are labelled by the spins $j_i$ and the group elements $g_i$ ($i=1,\cdots,6$). The action of the holonomy operator $\chi_{j}(g_4g_5g_6)$ translates, upon recoupling, into the functional depicted above, where intersections of links stand for $3jm$-symbols, and the spins $k_1,k_2,k_3$ are summed over. The dashed lines are the links carrying the spin $j$.
 \label{picturerechol} }
\end{figure}

Another way to derive recurrence relations on the 6j-symbols is to ``come back" to spin network functionals and look at the action of holonomy operators. Let us consider the spin network functional based on the tetrahedral graph labeled by representations $j_1,..,j_6$:
\be\label{spinnet}
\vphi_{\{j_l\}}(g_1,\dotsc,g_6)
\,=\,
\prod_{l=1}^6D^{(j_l)}_{a_lb_l}(g_l)\prod_{n=1}^4 i_n.
\ee
Here the intertwiners $i_v$ are given by the normalized Wigner's $3jm$-symbols, and its evaluation at the identity gives the 6j-symbol,
\be
\left.\vphi_{\{j_l\}}(g_1,\dotsc,g_6)\right|_{g_l=\id}
\,=\, \sixj{j_1& j_2 & j_3}{j_4& j_5 & j_6}.
\ee
The underlying tetrahedral graph is such that the edges carrying $j_4$, $j_5$ and $j_6$ form a triangle. Consider now the holonomy operator $\chi_{j}(g_4g_5g_6)$, which simply acts by multiplication on the spin network functional, $\chi_{j}(g_4g_5g_6)\,\vphi_{\{j_l\}}(g_1,..,g_6)$. We can recouple the matrix elements of $\chi_{j}(g_4g_5g_6)$ with the matrix elements of $g_4$, $g_5$ and $g_6$ already present in the functional $\vphi_{\{j_l\}}(g_1,..,g_6)$.
This gives a recurrence relation between spin network states,
\begin{multline} \label{rechol}
\chi_j(g_4g_5g_6)\vphi_{\{j_l\}}(g_1,\dotsc,g_6) = \sum_{k_4,k_5,k_6} (-1)^{j_1+j_2+j_3+j_4+j_5+j_6+k_4+k_5+k_6 + j} d_{k_4} d_{k_5} d_{k_6} \\
\times \begin{Bmatrix} k_4& j_4 & j \\ j_6& k_6 & j_2\end{Bmatrix}
\begin{Bmatrix}k_5& j_5 & j \\ j_4& k_4 & j_3 \end{Bmatrix}
\begin{Bmatrix}k_6& j_6 & j \\ j_5& k_5 & j_1\end{Bmatrix}
\vphi_{\{j_1,j_2,j_3,k_4,k_5,k_6\}}(g_1,\dotsc,g_6),
\end{multline}
which can be represented as in picture \ref{picturerechol}. The coefficients $d_k\equiv (2k+1)$ are the dimensions of the spin $k$ representations.

%
Evaluating both sides at $g_l=\id$, we find
\begin{multline} \label{rechol1}
(2j+1)\begin{Bmatrix}j_1& j_2 & j_3 \\ j_4& j_5 & j_6\end{Bmatrix} =
\sum_{k_4,k_5,k_6} (-1)^{j_1+j_2+j_3+j_4+j_5+j_6+k_4+k_5+k_6 + j} d_{k_4} d_{k_5} d_{k_6}
\\ \times \begin{Bmatrix} k_4& j_4 & j \\ j_6& k_6 & j_2\end{Bmatrix}
\begin{Bmatrix}k_5& j_5 & j \\ j_4& k_4 & j_3 \end{Bmatrix}
\begin{Bmatrix}k_6& j_6 & j \\ j_5& k_5 & j_1\end{Bmatrix}
\begin{Bmatrix}j_1& j_2 & j_3 \\ k_4& k_5 & k_6\end{Bmatrix}.
\end{multline}
Such recurrence relations for the 6j-symbol are well-known \cite{varsha}.
What is interesting about them is the fact that they can be derived by the action of holonomy operators, and thus can be directly extended to more general spin network states. Furthermore, the technique then gives recurrence relations to more general $\{3nj\}$-symbols, as we will discuss below in section \ref{fifj-recurrences}. In fact, we will see that recurrence relations for the $\{3nj\}$-symbols are naturally associated to cycles of the underlying graph, as one can guess from the above example.

But before moving to more general symbols, there is an interesting interpretation of \Ref{rechol1}.
Recall that the topological invariance of the Ponzano-Regge model is proved showing the invariance of the partition function under the two tridimensional Pachner moves, the 2-3 and the 1-4. Invariance under the 2-3 move is a direct consequence of the BE identity. Invariance under the 1-4 on the other hand is more subtle. Starting from BE and the orthogonality relation, one obtains indeed a formula formally matching one 6j symbol to four summed over four common spins:
\be \label{div14}
\Bigl(\sum_{l_4}d_{l_4}^2\Bigr)
\left\{\begin{array}{ccc} j_1 & j_2 & j_3 \\ j_4 & j_5 & j_6  \\ \end{array}\right\}
=
\sum_{l_1 \ldots l_4} (-)^{\sum_{i=1}^6 j_i} \bigg(\prod_{i=1}^4 (-)^{l_i} d_{l_i}\bigg)
\left\{\begin{array}{ccc} j_1 & j_2 & j_3 \\ l_1 & l_2 & l_3  \\ \end{array}\right\}
\left\{\begin{array}{ccc} j_6 & j_5 & j_1 \\ l_2 & l_3 & l_4  \\ \end{array}\right\}
\left\{\begin{array}{ccc} j_4 & j_2 & j_6 \\ l_3 & l_4 & l_1  \\ \end{array}\right\}
\left\{\begin{array}{ccc} j_3 & j_5 & j_4 \\ l_4 & l_1 & l_2  \\ \end{array}\right\}
\ee
which has the right geometric structure to represent the 1-4 move. However this identity is formal because both sides diverge. This is usually cured introducing a cut-off on the spins \cite{Ponzano}, and the resulting model is argued to be triangulation independent in the limit in which the cut-off is removed (see \cite{BarrettIleana} for a discussion). Notice then that the cut-offed 1-4 move can be reconstructed from \Ref{rechol1} simply multiplying by $d_j$ and summing over $j$ up to the desired cut-off. As we send the cut-off to infinity, the extra sum in the right hand side diverges, and the left hand side reproduces the correct diverging factor $\delta(\id) = \sum_j d_j^2$. The recurrence relation \Ref{rechol1} thus provides an alternative regularization of the 1-4 move, where instead of a cut-off on the spins, we fixed to the value $j$ one of the spins being summed over. This alternative regularization can be seen as a partial gauge-fixing in the spirit of \cite{PR1}, and we will use a similar method to gauge-fix the Pachner moves in 4d. Conversely, we may see \Ref{rechol1} as the equality of the summands of the above formal 1-4 move for any fixed spin $l_4$. In this sense, the recurrence relations coming from the 1-4 move are obtained by gauge-fixing the value of the spin responsible for the divergent factor.

Furthermore, the relation between \Ref{rechol1} and \Ref{div14} can be extended to spin network functionals. Indeed, instead of acting on the state $\vphi_{\{j_i\}}$ with a single character, let us act with the delta distribution $\delta(g_4 g_5 g_6)=\sum_j d_j \chi_j(g_4 g_5 g_6)$. The result can be easily obtained from \Ref{rechol}:
\begin{multline}
\delta\bigl(g_4 g_5 g_6\bigr)\,\vphi_{\{j_l\}}(g_1,\dotsc,g_6) = \sum_{k_4,k_5,k_6,j} (-1)^{j_1+j_2+j_3+j_4+j_5+j_6+k_4+k_5+k_6 + j} d_{k_4} d_{k_5} d_{k_6} d_j\\
\times \begin{Bmatrix} k_4& j_4 & j \\ j_6& k_6 & j_2\end{Bmatrix}
\begin{Bmatrix}k_5& j_5 & j \\ j_4& k_4 & j_3 \end{Bmatrix}
\begin{Bmatrix}k_6& j_6 & j \\ j_5& k_5 & j_1\end{Bmatrix}
\vphi_{\{j_1,j_2,j_3,k_4,k_5,k_6\}}(g_1,\dotsc,g_6),
\end{multline}
The evaluation of this equation at the identity reproduces the formal 1-4 move \Ref{div14}, in which the presence of the divergent factor $\delta(\id)$ now becomes obvious. In addition, notice that $\delta(g_4 g_5 g_6)$ imposes the classical flatness constraint $F=0$ on the triangle which carries $g_4$, $g_5$ and $g_6$. Therefore,
the 1-4 move is really generated through the action of the flatness constraint on a spin network state, a further connection between the classical symmetries and quantum recurrence relations which was already pointed out in \cite{ooguri3d}.


\section{Recurrence relations for $\{3nj\}$-symbols} \label{fifj-recurrences}

In the previous section we described three different methods to obtain recurrence relations for the 6j-symbols,
(i) inserting an existing constraint in the integral representation, (ii) using the defining BE identity, and (iii) acting with holonomy operators. With the exception of \Ref{recurrenceiso6j}, the relations we found are well-known in the literature. When moving to higher $3nj$-symbols the literature is rather scarce, with the notable exception of the 9j symbol, see \cite{varsha}. Because higher $3nj$-symbols are relevant in models of quantum gravity, it is useful to show how the techniques described earlier can be applied to obtain recurrence relations for general symbols.
In this and the next Sections we describe how one can use the techniques learned above to generate recurrence relations for general symbols.
Below in section \ref{BCrec} we will show an application of (i) to the 10j symbol.

Consider an arbitrary $\{3nj\}$-symbol, represented by a closed graph made of 3-valent vertices and whose links carry $\SU(2)$ irreducible representations, collectively denoted $\{j_i\}$. The evaluation of the graph can be obtained according to conventional rules assigning Clebsch-Gordan coefficients (or $3mj$-coefficients) to vertices. To unambiguously represent symbols with such spin network graphs, we use the conventions of \cite{varsha} for the orientation of links and vertices.

A straighforward way to obtain recurrence relations is to exploit the decomposition of the $3nj$-symbols into lower symbols.
This decomposition can be made isolating a part of the graph containing a cycle, as shown in Fig 2.
In the simplest case, the cycle contains only three links, and the symbol decomposes as
$\{3nj\} = \{6j\} \{3(n-1)j\}$. For a cycle with four links, $\{3nj\} = \sum_i \{9j\}_i \{3(n-1)j\}_i$, and so on.

We show how certain types of recurrence relations can be obtained using these decompositions and applying the known relations of the 6j and the $\{9j\}$. This procedure works quite well for decompositions built from 3-cycles and 4-cycles, but becomes cumbersome for larger cycles. Then, in the next Section we focus on $n=5$, the relevant case to 4d models of quantum gravity, and describe how these recurrence relations can be obtained from a regularized version of the 2-4 move. Interestingly, this derivation will open a window on new recurrence relations, which apply to graphs with larger cycles.

\begin{figure}
\begin{center}
\includegraphics[width=11cm]{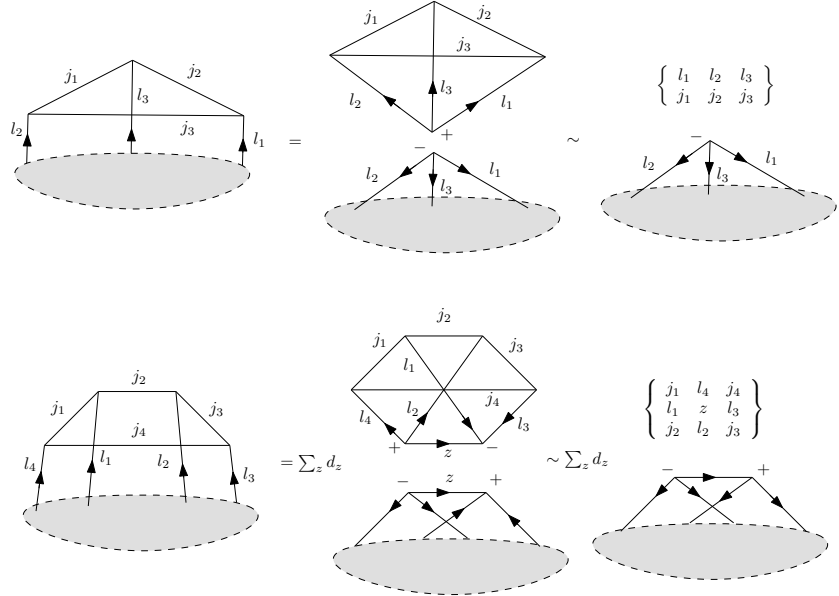}
\end{center}
\caption{ \label{reducible} The top figure displays a reducible symbol: the cycle $(j_1 j_2 j_3)$ can be factorized because there is an unique intertwiner from $l_1\otimes l_2\otimes l_3$ to $\C$. One can then use the recurrence relations of the 6j-symbol to shift the spins $(j_i)$. For a cycle made of four links, one needs to sum over the intertwiners $z$ between $l_4\otimes l_2\rightarrow l_1\otimes l_3$. One can then use recurrence relations which shift some $(j_i)$ provided they do not depend on $z$, but only on the spins which join the cycle.}
\end{figure}

\subsection{Recurrence relations from smaller symbols}
The simplest situation is a $\{3nj\}$ symbol in which we can isolate a cycle formed by three links, see top panel of Figure
\ref{reducible}. In this case the symbol is reducible, as it can be written as a smaller $\{3(n-1)j\}$-symbol times a 6j-symbol. Since the labels of the 6j-symbol do not enter the larger symbol, we can trivially use the recurrence relations of the 6j to infer relations on the $\{3nj\}$ symbol, for instance
\be
A_{+1}(j)\begin{Bmatrix} l_1 &l_2 &l_3 \\ j_1+1 &j_2 &j_3 \\ \hdotsfor[2]{3}\end{Bmatrix} +
A_0(j)\begin{Bmatrix} l_1 &l_2 &l_3 \\ j_1 &j_2 &j_3 \\ \hdotsfor[2]{3}\end{Bmatrix} +
A_{-1}(j)\begin{Bmatrix} l_1 &l_2 &l_3 \\ j_1-1 &j_2 &j_3 \\ \hdotsfor[2]{3}\end{Bmatrix} = 0.
\ee

Consider next a symbol containing a cycle made of four links, bottom panel of Figure
\ref{reducible}. We can still apply the recoupling theory to write the symbol in terms of smaller ones, but this time a summation over smaller symbols appears: the original symbol is not reducible this way.
This does not stop us from applying recurrence relations to the smaller symbols, but care is needed to avoid including into them the spin being summed over. Using the example in the figure, one can derive relations for the symbol to the left starting from relations for the \nj-symbol on the right which do not include the spin $z$ intertwining the representations $\{l_i\}$ together. For instance,
\begin{multline}
\Bigl[\f{(j_2+l_2+j_3+1)(j_2+j_3-l_2)(-j_4+j_3+l_3)(j_4+l_3-j_3+1)}{(j_1+l_1+j_2+2)(j_1-l_1+j_2+1)}\Bigr]^{\f{1}{2}} \begin{Bmatrix} j_1+\f{1}{2} &l_4 &j_4+\f{1}{2} \\ l_1 &z &l_3 \\ j_2-\f{1}{2} &l_2 &j_3-\f{1}{2} \\ \hdotsfor[2]{3} \end{Bmatrix} \\
+ \Bigl[\f{(j_2-j_3+l_2)(-j_2+j_3+l_2+1)(j_4+j_3+l_3+2)(j_4+j_3-l_3+1)}{(j_1+l_1+j_2+2)(j_1-l_1+j_2+1)}\Bigr]^{\f{1}{2}}\begin{Bmatrix} j_1+\f{1}{2} &l_4 &j_4+\f{1}{2} \\ l_1 &z &l_3 \\ j_2-\f{1}{2} &l_2 &j_3+\f{1}{2} \\ \hdotsfor[2]{3} \end{Bmatrix} \\
+
\Bigl[\f{(j_2+l_2+j_3+2)(j_2+j_3-l_2+1)(j_4+j_3+l_3+2)(j_4+j_3-l_3+1)}{(j_1+l_1-j_2+1)(-j_1+l_1+j_2)}\Bigr]^{\f{1}{2}}\begin{Bmatrix} j_1+\f{1}{2} &l_4 &j_4+\f{1}{2} \\ l_1 &z &l_3 \\ j_2+\f{1}{2} &l_2 &j_3+\f{1}{2} \\ \hdotsfor[2]{3} \end{Bmatrix} \\
-
\Bigl[\f{(j_2-j_3+l_2+1)(-j_2+j_3+l_2)(-j_4+j_3+l_3)(j_4+l_3-j_3+1)}{(j_1+l_1-j_2+1)(-j_1+l_1+j_2)}\Bigr]^{\f{1}{2}}\begin{Bmatrix} j_1+\f{1}{2} &l_4 &j_4+\f{1}{2} \\ l_1 &z &l_3 \\ j_2+\f{1}{2} &l_2 &j_3-\f{1}{2} \\ \hdotsfor[2]{3} \end{Bmatrix} \\
= (2j_2+1)(2j_3+1)\Bigl[\f{(j_1+l_4+j_4+2)(j_1+l_4-j_4+1)}{(j_1+l_1-j_2+1)(-j_1+l_1+j_2)(j_1+l_1+j_2+2)(j_1-l_1+j_2+1)}\Bigr]^{\f{1}{2}}\begin{Bmatrix} j_1 &l_4 &j_4 \\ l_1 &z &l_3 \\ j_2 &l_2 &j_3 \\ \hdotsfor[2]{3} \end{Bmatrix}
\end{multline}
Other relations can be obtained from other relations of the \nj-symbol given in \cite{varsha}.

Notice that the coefficients of recurrence relations obtained this way will depend on the spins of the cycle and on those of the links directly joining the cycle.
This property also holds for cycles made of five and six links. However the procedure can become quite cumbersome for such cycles. Below we will derive an alternative which is more fruitful for large cycles.

Let us also mention an alternative used by Jang \cite{jang}. A $\{3(n+1)j\}$-symbol can be expressed in terms of a sum of products of a $\{3nj\}$-symbol and typically some 6j-symbols. Then, using symmetries of the $\{3(n+1)j\}$-symbol, one can get non-trivial relations between the $\{3nj\}$-symbols, with some 6j as coefficients. However, it seems that this method works for symbols of the first kind but may not be efficient for other kinds of symbols. Moreover, it does not provide any geometric picture such as simplicial moves, in contrast to the method we will use.

\section{The 15j symbol and 4d BF theory}

Let us now fix $n=5$. The resulting 15j symbol is not unique, unlike the 6j and 9j symbols. Using the terminology of \cite{jucys, varsha}, one can distinguish five different irreducible kinds.
Depending on the kind, different types of cycles will be present, and one can derive relations for the 15j symbol proceeding as described above. All links of the 15j-symbols of the first and second kinds can be included in cycles made of four links, but five links are necessary for the 15j of the fifth kind. The symbols of the third and fourth kinds contain both types of cycles.

The 15j symbol was used by Ooguri \cite{ooguri} as the vertex amplitude of four-dimensional BF theory with gauge group SU(2), a theory often used as a starting point for spin foam models of quantum gravity. To see how the 15 symbol enters the game, let us recall that Ooguri's model is built triangulating spacetime with a collection of flat 4-simplices glued together, and assigning an $\SU(2)$ element $g_t$ to each tetrahedron $t$. Within the triangulation, each triangle $f$ will be shared by many tetrahedra, and one can define the quantity $g_f = \prod_{t \supset f} g_t$, called the holonomy associated to each $f$. Finally, the partition function of the model is the integral over all the $g_t$ of $\prod_f\delta(g_f)$. These delta functions can be seen as imposing a discretized version of the flatness equation which defines the topological BF field theory.

The integrals can be evaluated using the Plancherel formula $\delta(g_f)=\sum_{j\in\f{\N}{2}}d_j\chi_j(g_f)$ at each triangle and then writing $\chi_j(g_f)=\chi_j(\prod_{f \supset f} g_t)$ as the trace of a product of matrices, one for each tetrahedron sharing the triangle.
Noticing that each tetrahedron contains four triangles, there is only four matrices containing a given $g_t$ in the partition function, which then factorizes into the following integrals,
\be \label{intg4}
\int_{\SU(2)}dg_t\ \bigotimes_{f=1}^4 D^{(j_f)}(g_t) = \mathrm{id}_{\mathrm{Inv}}.
\ee
Here $\mathrm{id}_{\mathrm{Inv}}$ is the identity on the invariant subspace of the tensor product of the four representations $j_f$ meeting at $t$. An orthonormal basis of such 4-valent intertwiners is given by a tree expansion, i.e. an expansion onto 3-valent intertwiners: one has to choose one of the three possible pairings between the four $j_f$, and sum over the internal representation. For instance, an intertwiner is $\lvert (j_1j_2),(j_3j_4);i\rangle$ where $i$ satisfies: $\lvert j_1-j_2\rvert\leq i\leq j_1+j_2$ and $\lvert j_3-j_4\rvert\leq i\leq j_3+j_4$. The identity is given by:
\be \label{pairing}
\mathrm{id}_{\mathrm{Inv}} = \sum_i\ \lvert (j_1j_2),(j_3j_4);i\rangle\ \langle (j_1j_2),(j_3j_4);i\rvert
\ee
The contractions of such intertwiners at each 4-simplex give rise to a 15j-symbol.
With respect to the original triangulation, the model naturally assigns a spin $j_f$ to each triangle, an intertwiner $i_t$ to each tetrahedron, and the resulting 15j symbol as the amplitude of a 4-simplex.

The attentive reader will have noticed that which 15-symbol emerges depends on the choice of pairing in \Ref{pairing} for each of the five intertwiners in the 4-simplex.
It is possible to write the model in terms of a single kind of 15j-symbol, but generally this requires using different pairings for each of the two intertwiners entering \eqref{pairing}. Gauge invariance is guaranteed by the subsequent emergence of appropriate 6j-symbols at the tetrahedra recoupling the two different pairings.\footnote{This is often overviewed in the literature on spin foam models, although these 6j-symbols are part of the model originally defined by Ooguri.}

A natural question is whether exploring all the possible pairings one can reconstruct all the five different kinds of 15 symbols. It is easy to show that one can indeed recover the 15j of the first, third, fourth and fifth kinds (in the terminology of \cite{jucys}), along with reducible symbols. On the other hand, we were not able to prove that the second kind can also be obtained in this way. We do not know whether 15j-symbols of the second kind can appear in the spin foam model.

The interest in Ooguri's model model is its topological invariance, which generalizes to four dimensions the analogue property of the Ponzano-Regge model. The proof uses again Pachner moves \cite{ooguri,alex}. In 4d, there are three independent moves, which are discussed in details in \cite{carter}: 3-3, 2-4 and 1-5. When applied to the vertex amplitude of the model, the 15 symbol, both the second and the third moves are divergent  \cite{carter}. As for the 1-4 move of the 6j, the source of the divergences arises from redundant delta functions imposing $\SU(2)$ flatness in the bulk.


Let us now come back to the recurrence relations that we can construct using the strategy of the previous section.
Since the 15j symbol gives rise to a topological model, one might wonder whether these relations are again a consequence of a more fundamental identity, like the BE identity for the 6j, and whether like the BE identity, they are related to a Pachner move.
We now show that this is indeed the case: the recurrence relations are a consequence of the (regularized) 2-4 move.
Since the 2-4 move is a component of the invariance under homeomorphisms of the model, the recurrence relations can be seen as difference equations contributing to the implementation of this symmetry, i.e. as part of a discrete version of the classical constraints of BF theory.


However the relevant move is divergent, so we need to regularize it first. We do so in the the subsection, and then move to the derivation of recurrence relations from it.

\subsection{Regularizing the 2-4 move}



\begin{figure}
\begin{center}
\includegraphics[width=5cm]{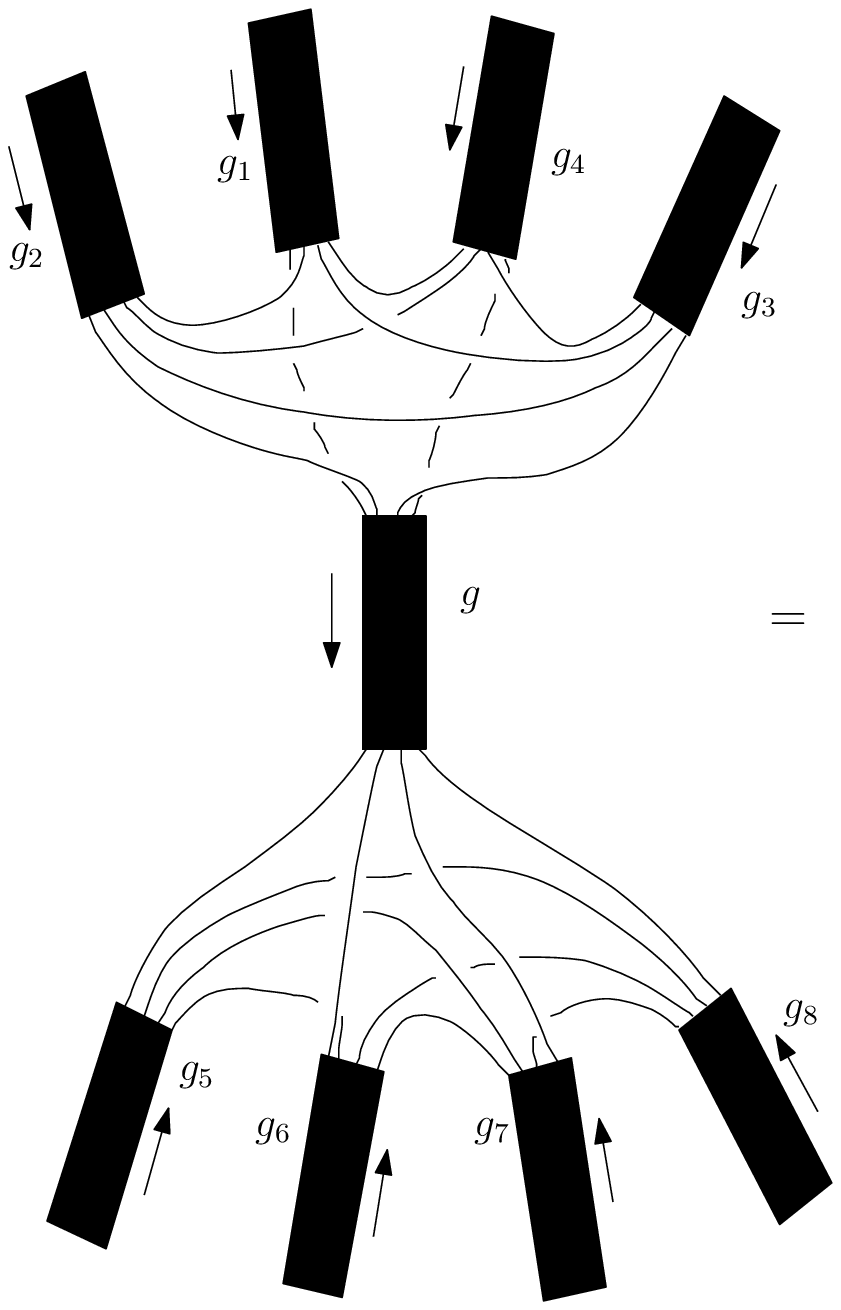}\qquad\qquad\includegraphics[width=8cm]{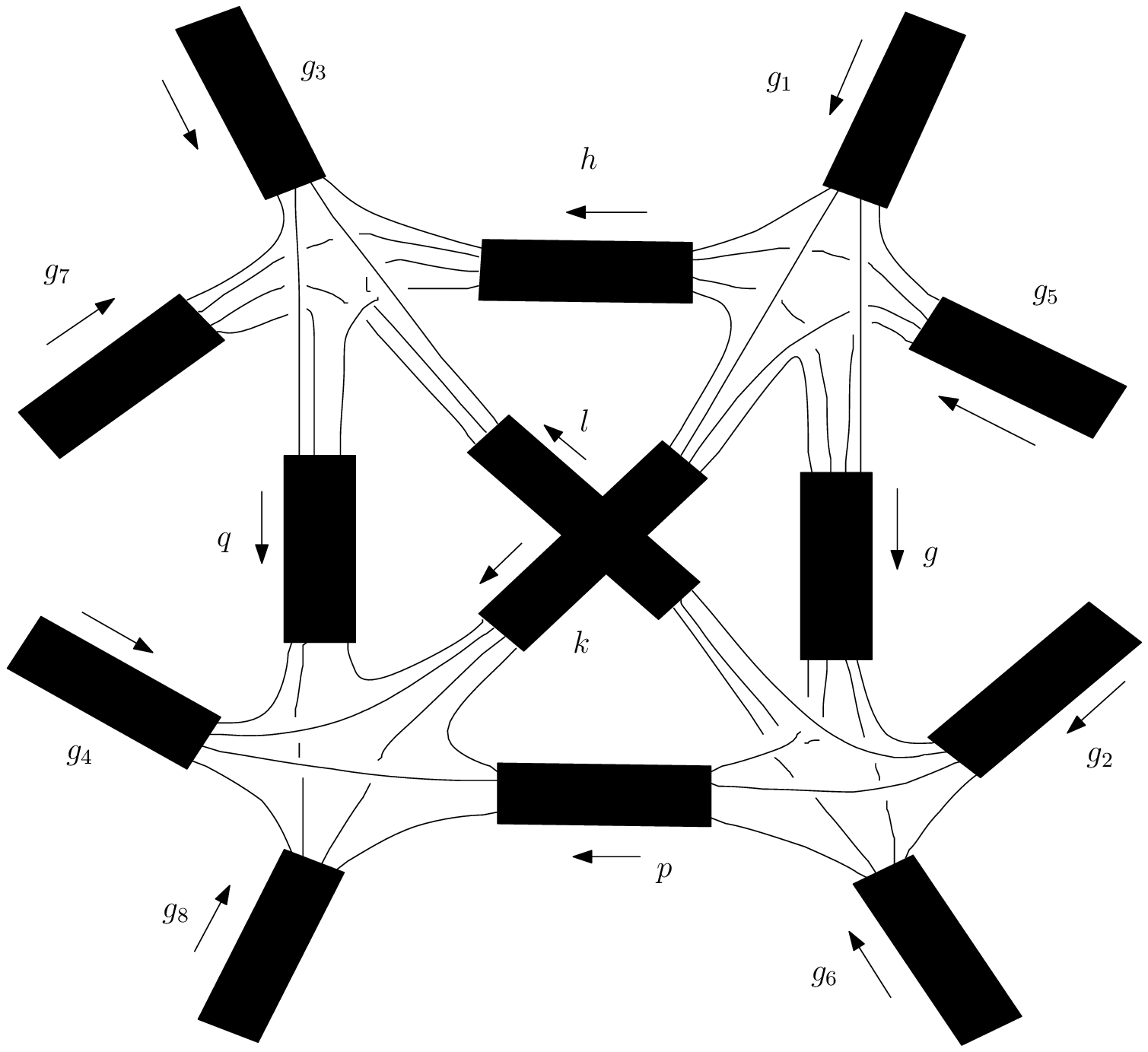}
\end{center}
\caption{ \label{move4-2gft} The left picture describes the configuration with two 4-simplices, while the right picture displays the structure of four 4-simplices which are glued to each other along different tetrahedra. The arrows give the orientations of the dual edges and the boxes stand for integration of the group elements carried by these dual edges. Once the right hand side is regularised, the two situations have the same BF amplitude. To regularise it, we have dropped the flatness condition for the face which passes through $l$, $q$ and $p$. As a consequence, one obtains only one 15j-symbol on the right hand side, since only three lines pass through $l$, $q$ and $p$: three 4-simplex amplitudes are reduced to $\{12j\}$-symbols. }
\end{figure}

Since the model is intrinsically defined without choices of pairings, but rather in terms of group integrations, it is convenient to consider the move in the group picture, see figure \ref{move4-2gft}.
Black boxes stand both for edges dual to tetrahedra and integrations over the corresponding group elements $g_t$, while lines denote the boundaries of the faces dual to triangles. Each box is crossed by four lines. On the two sides of the move, there are eight boundary tetrahedra, with elements $(g_i)_{i=1,\dots,8}$. On the left side, the two 4-simplices are glued along a single tetrahedron, with element $g$.
Using the orientations of dual edges given in figure \ref{move4-2gft}, the amplitude of the left hand side is
\begin{multline} \label{move4-2lhs}
\int \prod_{t=1}^8 dg_t\,D^{(j_{12})}(g_1 g_2\mone)\,D^{(j_{13})}(g_1 g_3\mone)\,D^{(j_{14})}(g_1 g_4\mone)\,D^{(j_{23})}(g_2 g_3\mone)\,D^{(j_{23})}(g_2 g_3\mone)\,D^{(j_{24})}(g_2 g_4\mone)\,D^{(j_{34})}(g_3 g_4\mone)\\
D^{(j_{56})}(g_5 g_6\mone)\,D^{(j_{57})}(g_5 g_7\mone)\,D^{(j_{58})}(g_5 g_8\mone)\,D^{(j_{67})}(g_6 g_7\mone)\,D^{(j_{68})}(g_6 g_8\mone)\,D^{(j_{78})}(g_7 g_8\mone)\\
\int dg\,D^{(j_{15})}(g_1\,g\, g_5\mone)\,D^{(j_{26})}(g_2\,g\, g_6\mone)\,D^{(j_{37})}(g_3\,g\, g_7\mone)\,D^{(j_{48})}(g_4\,g\, g_8\mone),
\end{multline}
where tensor products are intended between the matrices $D^{(j)}$. After integration over the elements $g_t$ for the eight boundary tetrahedra, one can contract the free indices of intertwiners, corresponding to the free ends of the open lines, with eight chosen intertwiners. This process includes choosing pairings of the virtual spins.
The integration over $g$, i.e. over the tetrahedron shared by the two 4-simplices, produces the sum \Ref{pairing}, and this leads precisely to the left picture of figure \ref{move4-2graph}.

On the right hand side, one has to integrate over six elements corresponding to the six bulk tetrahedra. Open lines correspond to boundary triangles, closed lines correspond to triangles in the bulk and are shared by exactly three tetrahedra (and three 4-simplices). Each bulk triangle contributes a factor $\delta(\prod_{t \supset f} g_t )$ to the amplitude. By inspection, one of this delta is redundant, and this causes the $\delta(\id)$ divergence, exactly like in the 1-4 move of 3d gravity. This can be removed brutally dropping one of the four flatness conditions, say
 $\delta(lqp\mone)$, since the constraint $lqp\mone=\id$ is ensured by the other delta functions. The resulting \emph{finite} amplitude is
\begin{multline}
\int \prod_{t=1}^8 dg_t\,D^{(j_{15})}(g_1 g_5\mone)\,D^{(j_{26})}(g_2 g_6\mone)\,D^{(j_{37})}(g_3g_7\mone)\,D^{(j_{48})}(g_4 g_8\mone)\ \int dg\,dh\,dk\,dl\,dp\,dq\\
\delta\bigl(g\,l\,h\mone\bigr)\,\delta\bigl(g\,p\,k\mone\bigr)\,\delta\bigl(h\,q\,k\mone\bigr)\ D^{(j_{12})}(g_1\,g\, g_2\mone)\,D^{(j_{56})}(g_5\,g\, g_6\mone)\ D^{(j_{57})}(g_5\,h\, g_7\mone)\,D^{(j_{13})}(g_1\,h\, g_3\mone)\\
D^{(j_{58})}(g_5\,k\, g_8\mone)\,D^{(j_{14})}(g_1\,k\, g_4\mone)\,D^{(j_{23})}(g_2\,l\, g_3\mone)\,D^{(j_{67})}(g_6\,l\, g_7\mone)\\
D^{(j_{24})}(g_2\,p\, g_4\mone)\,D^{(j_{68})}(g_6\,p\, g_8\mone)\,D^{(j_{34})}(g_3\,q\, g_4\mone)\,D^{(j_{78})}(g_7\,q\, g_8\mone).
\end{multline}
The bulk integrals can be performed as usual, expanding the delta functions with the Plancherel formula and using repeatedly \eqref{intg4}. Once again, the latter step introduces a choice of pairing. Notice however that the result will involve only one 15j symbol: because a delta function has been removed on one triangle, the group elements $l$, $p$ and $q$ now appear only three times. Thus the amplitude for the three 4-simplices sharing this triangle are reduced to smaller 12j-symbols.

The two expressions coincide after the regularization, since all we have changed was to remove a redundant delta function. It can also be explicitly proved using the graphical method of \cite{alex}. The scheme is basically as follows. Notice first that $g$, on the left hand side, can be reabsorbed on the right of $g_1$, $g_2$, $g_3$ and $g_4$, using the translation invariance of the Haar measures. We briefly describe how the right hand side reduces to the same amplitude. The same method can be used to eliminate $h$ and $q$ for instance, so that $k$ is forced to be the identity. $g$ can also be reabsorbed, enforcing $l=\mathrm{id}$. Finally, integrating $p$ is trivial because one gets from the previous simplifications $\delta(p)$.

Because the divergence is exactly the same as in the 3d 1-4 move, we can obtain formulas similar to \Ref{rechol1}. Indeed, instead of removing the delta function $\delta(lqp\mone)$, let us keep a component, say of spin $J$, of its Fourier expansion, $\chi_J(lqp\mone)$ in the integrand. Performing the integrals and choosing the pairings, the right hand side of the move is now made of four $15j$s, with a dependence on $J$. Let us see how it changes the amplitude of the left hand side. Since the other delta functions (in the right hand side) enforce $lqp\mone=\id$, the character $\chi_J(lqp\mone)$ can be reduced to $d_J$ which comes in factor of the above expression. Thus, the two $15j$s of the left hand side are only changed by a factor $d_J$, like in \Ref{rechol1}.


We now show how this regularized move can be used to obtain the previous recurrences and new ones.

\subsection{Recurrence relations from the 2-4 move} \label{rec2-4move}

\begin{figure}
\begin{center}
\includegraphics[width=10cm]{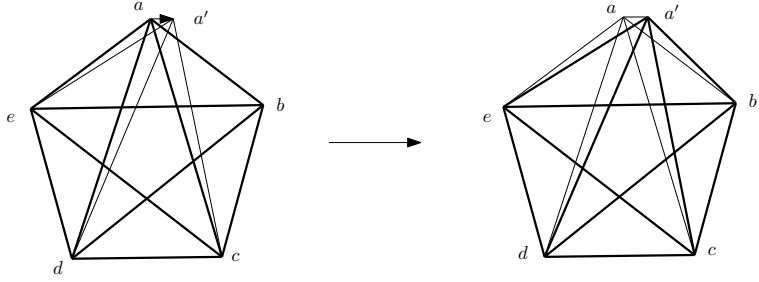}
\end{center}
\caption{ \label{move4-2} The two sides of the 2-4 move. On the left hand side, the 4-simplices $(abcde)$ and $(aa'cde)$ are glued along the tetrahedron $(acde)$. On the right hand side the four 4-simplices $(a'bcde)$, $(a'bacd)$, $(a'bade)$ and $(a'bace)$. This configuration has four 'bulk' triangles (given by $(a'bc)$, $(a'bd)$, $(a'be)$ and $(aa'b)$), which are shared by three 4-simplices, and six 'bulk' tetrahedra (made of the points $a'$ and $b$ together with any choice of two others), which are shared by two 4-simplices. When the spins of the triangles $(aa'e)$ and $(aa'c)$ are taken to be zero, we may consider $a'$ as being  very close to $a$. Then, the areas of the initial 4-simplex $(abcde)$ on the left and those of $(a'bcde)$ on the right only differ by some slight shifts for the three triangles sharing the edge $(ad)$.
}
\end{figure}

Following the derivation of \Ref{recurrenceiso6j} from the BE identity, we look at the regularized 2-4 move with some of the boundary spins fixed. The choice has to be careful, to ensure that we still have one 15j symbol on the right hand side. The situation is illustrated by Figure \ref{move4-2}: we consider a 4-simplex, with points $a,b,c,d,e$, and we glue another one, denoted $(aa'cde)$, along the tetrahedron $(acde)$. This will be the left hand side of the 2-4 move. Every triangle is colored with a fixed $\SU(2)$ representation, and every tetrahedron, except $(acde)$, by an fixed intertwiner: these are the 'external', or boundary, data. The tetrahedron $(acde)$ being shared by the two 4-simplices, the corresponding intertwiner is summed over to glue them (due to the integral over $g$ in \Ref{move4-2lhs}). The right hand side configuration is made of four 4-simplices respectively obtained by dropping the points $a,c,d$ and $e$.

\begin{figure}
\begin{center}
\includegraphics[width=17cm]{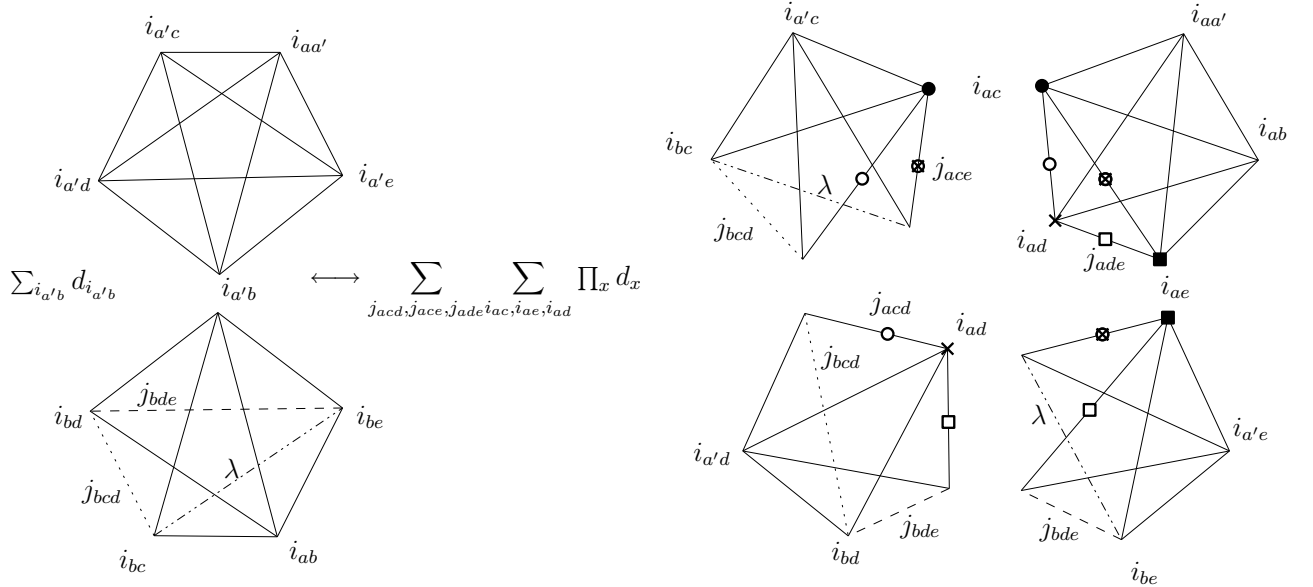}
\end{center}
\caption{ \label{move4-2graph} The regularized 4-2 Pachner move. Only one graph on the right hand side is a 15j-symbol since the flatness condition for the triangle $(aa'b)$ has been dropped. $x$ stands for the summed colors, whose corresponding links and vertices (intertwiners are not displayed) are marked. To get formulas involving only one 15j-symbol on each side of the move, we take $j_{bcd}=j_{bde}=0$, which are represented by dotted and dashed lines, so that many graphs reduce to (products of) 6j-symbols. Looking at small values of $\lambda\equiv j_{bce}$ enables to get interesting explicit recurrences.}
\end{figure}

To write the indices of spins and intertwiners, we will use the standard dual notation: we denote a simplex label by only writing the points which are {\it not} vertices of the simplex. Typically, the representation coloring the triangle made of the points $(cde)$ is $j_{aa'b}$ and that coloring the tetrahedron $(bcde)$ is $i_{aa'}$. To regularize the move, we need to drop the delta function of one internal face of the right hand side: we remove that of the triangle $(aa'b)$. Thus, only the amplitude of the 4-simplex which do not share this triangle, i.e. $(a'bcde)$, is actually a 15j, while those of the other three 4-simplices are reduced to some smaller 12j-symbols, as indicated in Figure \ref{move4-2graph}.

We now fix two boundary spins to zero: $j_{bcd}=j_{bde}=0$. The corresponding links in the spin network graphs are represented by dotted and dashed lines in Figure \ref{move4-2graph}. This enforces the equality $j_{abd}=j_{a'bd}$ between boundary spins. As a consequence, the 15j of $(aa'cde)$ on the left hand side turns into a (possibly reducible) 9j-symbol. It turns out that for any pairing of the boundary intertwiner $i_{ab}$, we can always expand the summed intertwiner $i_{a'b}$ so that this 9j is further reduced to a 6j-symbol or at least a product of two 6j-symbols. This way, the left hand side contains only one 15j and one or two 6js.

A similar reduction applies to the right hand side. In particular, among the spins which are summed over, two become fixed: $j_{acd}=j_{a'cd}$ and $j_{ade}=j_{a'de}$. Also, the three 12j-symbols are reduced to products of 6j-symbols, whose precise forms depend on the choice of pairings for the boundary intertwiners.

We are then in a position to write relations with one 15j on each side, and with coefficients being 6j-symbols. It is also easy to see that all these 6j-symbols depend on the spin $\lambda\equiv j_{bce}$. This spin also parametrizes the ranges of the data which are summed: typically, one gets from Figure \ref{move4-2graph} that $\lvert j_{a'ce}-\lambda\rvert\leq j_{ace}\leq j_{a'ce}+\lambda$. Thus, exactly as we specialize one spin to $1/2$ in the BE identity to get recurrence relations (or to 1, to get recurrences with shifts of 1), we can specialize $\lambda$ to small values and this way obtain recurrence relations. Notice also that the conditions $j_{bcd}=j_{bde}=0$ entail some relations between the boundary spins, which also depend on $\lambda$, namely: $j_{abc}=j_{a'bc}+\alpha$ and $j_{abe}=j_{a'be}+\beta$, where $\alpha$ and $\beta$ are such that: $-\lambda\leq \alpha,\beta\leq\lambda$.

Before we give the general formulas, we would like to have a geometric picture of what we have done with the 2-4 move. In 3d, the geometric content of the move and of the recurrence relations in terms of gluing some flattened tetrahedra is supported by the interpretation of spins as quantum lengths.
In 4d, a natural interpretation of the spins labelling triangles and of the intertwiners labelling tetrahedra comes from the geometric quantization of a single tetrahedron and the similarity of BF theory and loop quantum gravity at the kinematical level. It invites us to think of the spins $j$ as being related to the triangle areas, and the intertwiners $i$ as areas of parallelograms chopping each tetrahedron in half (notice that the three possible pairings of each intertwiner correspond to the different ways of chopping each tetrahedron in half) \cite{baez}.

Then, choosing $j_{bcd}=j_{bde}=0$ means that the triangles $(aa'e)$ and $(aa'c)$ are ''very small``, so that the 4-simplex $(aa'cde)$ is flattened. From the point of view of the 4-simplex $(abcde)$, the overall result can be seen as a small ''displacement`` of the point $a$ to $a'$, see Figure \ref{move4-2}. On the one hand, the relations $j_{acd}=j_{a'cd}$ and $j_{ade}=j_{a'de}$ mean that the new triangles $(a'be)$ and $(a'bc)$ have the same areas as the initial triangles $(abe)$ and $(abc)$. On the other hand, the areas of the triangles sharing the edge $(ad)$ are shifted when moving the point $a$: as we have seen, $j_{abc}=j_{a'bc}+\alpha$ and $j_{abe}=j_{a'be}+\beta$, and $j_{a'ce}$ is also shifted by a quantity depending on $\lambda$. Varying $\lambda$ corresponds to moving the point $a'$ so that the area of the triangle $(aa'd)$ changes, while keeping those of $(aa'c)$ and $(aa'e)$ fixed and very small.

Whether or not the intertwiners are shifted when moving $a$ to $a'$ (for instance, whether or not $i_{ab}=i_{a'b}$) depends on the chosen pairings. More precisely, we look at the tree expansion of an intertwiner in the spin network picture. If two shifted spins meet at a common vertex of the intertwiner, then the spin $i$ of the latter will not be changed. But if the line carrying the spin $i$ is inserted between two shifted spins, it will also be shifted. This way, one can obtain recurrence relations for 15j-symbols whose cycles have different numbers of links, by changing the pairings of intertwiners.

Relations for cycles made of three and four links turn out to be precisely those for 6j and \nj-symbols, as expected. Let us present new relations, adapted to cycles of five and six links. There are two ways of inserting intertwiners between the shifted links to get a cycle of five links. Indeed, one can insert, or not, $i_{a'b}$ between $j_{a'bc}$ and $j_{a'be}$ on the left hand side, while if $i_{a'b}$ is inserted, it is then equivalent to insert $i_{a'e}$ or $i_{a'c}$. We thus get two different relations:
\begin{multline} \label{5-cycle 1}
(-1)^{R}\sum_{i_{a'b}=\lvert i_{ab}-\lambda\rvert}^{i_{ab}+\lambda} d_{i_{a'b}} \begin{Bmatrix} \lambda &i_{a'b} &i_{ab} \\ j_{a'bd} &j_{a'bc}+\alpha &j_{a'bc} \end{Bmatrix} \begin{Bmatrix} \lambda &i_{a'b} &i_{ab} \\ j_{aa'b} &j_{a'be}+\beta &j_{a'be} \end{Bmatrix} \begin{Bmatrix} i_{a'b} &{} &i_{a'e} &{} &j_{a'bc} \\ {} &j_{a'be} &{} &j_{a'ce} &{} \\ \hdotsfor[2]{5}\end{Bmatrix} \\
= \sum_{j_{ace},i_{ae}} d_{i_{ae}}d_{j_{ace}} \begin{Bmatrix} i_{a'c} &j_{a'bc} &j_{a'ce} \\ \lambda &j_{ace} &j_{a'bc}+\alpha \end{Bmatrix} \begin{Bmatrix} \lambda &i_{a'e} &i_{ae} \\ j_{a'de} &j_{abe}+\beta &j_{a'be} \end{Bmatrix} \begin{Bmatrix} \lambda &i_{a'e} &i_{ae} \\ j_{aa'e} &j_{ace} &j_{a'ce} \end{Bmatrix} \begin{Bmatrix} i_{ab} &{} &i_{ae} &{} &j_{a'bc}+\alpha \\ {} &j_{a'be}+\beta &{} &j_{ace} &{} \\ \hdotsfor[2]{5}\end{Bmatrix}
\end{multline}
where $\alpha$ and $\beta$ are half-integers between $-\lambda$ and $\lambda$, $d_j\equiv 2j+1$ and $R=j_{a'de}+j_{a'bd}-j_{aa'e}-j_{aa'b}-i_{ac}+2j_{a'bc}+\alpha+\lambda$;
\begin{multline} \label{5-cycle 2}
\begin{Bmatrix} i_{ab} &j_{a'bc} &j_{a'be} \\ \lambda &j_{a'be}+\beta &j_{a'bc}+\alpha \end{Bmatrix} \begin{Bmatrix} j_{a'bc} &{} &i_{a'e} &{} &i_{a'c} \\ {} &j_{a'be} &{} &j_{a'ce} &{} \\ \hdotsfor[2]{5}\end{Bmatrix} = \sum_{i_{ae},j_{ace},i_{ac}} (-1)^S d_{i_{ae}}d_{j_{ace}}d_{i_{ac}} \begin{Bmatrix} \lambda &i_{a'e} &i_{ae} \\ j_{a'de} &j_{a'be}+\beta &j_{a'be} \end{Bmatrix} \\
\times \begin{Bmatrix} \lambda &i_{a'e} &i_{ae} \\ j_{aa'e} &j_{ace} &j_{a'ce} \end{Bmatrix} \begin{Bmatrix} \lambda &i_{a'c} &i_{ac} \\ j_{a'cd} &j_{a'bc}+\alpha &j_{a'bc} \end{Bmatrix} \begin{Bmatrix} \lambda &i_{a'c} &i_{ac} \\ j_{aa'c} &j_{ace} &j_{a'ce} \end{Bmatrix} \begin{Bmatrix} j_{a'bc}+\alpha &{} &i_{ae} &{} &i_{ac} \\ {} &j_{a'be}+\beta &{} &j_{ace} &{} \\ \hdotsfor[2]{5}\end{Bmatrix}
\end{multline}
for $S= i_{ab}+i_{ac}-i_{a'c}+2i_{a'e}+2j_{a'bc}+\alpha+j_{ade}+j_{aa'e}+j_{acd}+j_{aa'c}$.

\begin{figure}
\begin{center}
\includegraphics[width=14cm]{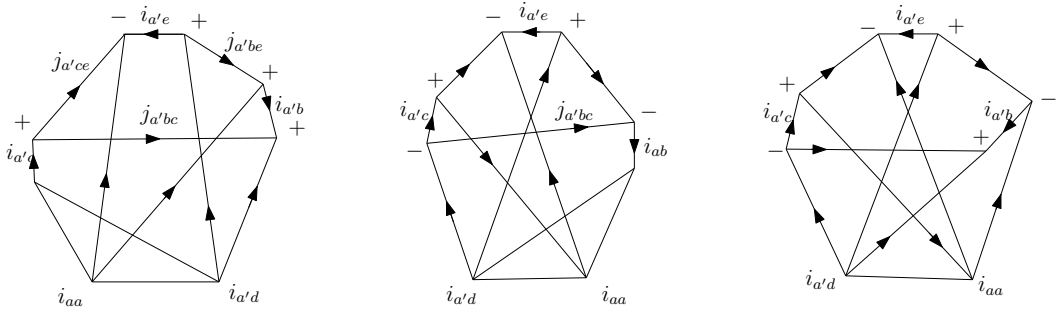}
\end{center}
\caption{ \label{fifj-move4-2} This picture represents the 15j-symbols respectively appearing in \eqref{5-cycle 1}, \eqref{5-cycle 2}, which give recurrences for a cycle of five links, and in \eqref{6-cycle} with a cycle of six links. Notice that only the orientations of the vertices and links of the cycles and of the links joining them are required.}
\end{figure}

There is only one way to make a cycle of six links, which gives the following relation:
\begin{multline} \label{6-cycle}
\sum_{i_{a'b}=\lvert i_{ab}-\lambda\rvert}^{i_{ab}+\lambda} d_{i_{a'b}} \begin{Bmatrix} \lambda &i_{a'b} &i_{ab} \\ j_{a'bd} &j_{a'bc}+\alpha &j_{a'bc} \end{Bmatrix} \begin{Bmatrix} \lambda &i_{a'b} &i_{ab} \\ j_{aa'b} &j_{a'be}+\beta &j_{a'be} \end{Bmatrix} \begin{Bmatrix} i_{a'b} &{} &i_{a'e} &{} &i_{a'c} &{} \\ {} &j_{a'be} &{} &j_{a'ce} &{} &j_{a'bc} \\ \hdotsfor[2]{6}\end{Bmatrix} \\
= \sum_{i_{ae},j_{ace},i_{ac}} (-1)^T d_{i_{ae}}d_{j_{ace}}d_{i_{ac}} \begin{Bmatrix} \lambda &i_{a'e} &i_{ae} \\ j_{a'de} &j_{a'be}+\beta &j_{a'be} \end{Bmatrix}
\begin{Bmatrix} \lambda &i_{a'e} &i_{ae} \\ j_{aa'e} &j_{ace} &j_{a'ce} \end{Bmatrix} \\ \times \begin{Bmatrix} \lambda &i_{a'c} &i_{ac} \\ j_{a'cd} &j_{a'bc}+\alpha &j_{a'bc} \end{Bmatrix} \begin{Bmatrix} \lambda &i_{a'c} &i_{ac} \\ j_{aa'c} &j_{ace} &j_{a'ce} \end{Bmatrix} \begin{Bmatrix} i_{ab} &{} &i_{ae} &{} &i_{ac} &{} \\ {} &j_{a'be}+\beta &{} &j_{ace} &{} &j_{a'bc}+\alpha \\ \hdotsfor[2]{6}\end{Bmatrix}
\end{multline}
with $T= 2j_{a'bc}+2\lambda+\alpha+i_{a'c}+i_{ac}+j_{a'bd}+j_{aa'b}+j_{aa'c}+j_{aa'e}+j_{a'de}+j_{a'cd}$. The coefficients of this equation are naturally those of the left hand side of \eqref{5-cycle 1} with those of the right hand side of \eqref{5-cycle 2}. The 15j-symbols described by \eqref{5-cycle 1}, \eqref{5-cycle 2} and \eqref{6-cycle} are represented in figure \ref{fifj-move4-2}.

\begin{figure}
\begin{center}
\includegraphics[width=17cm]{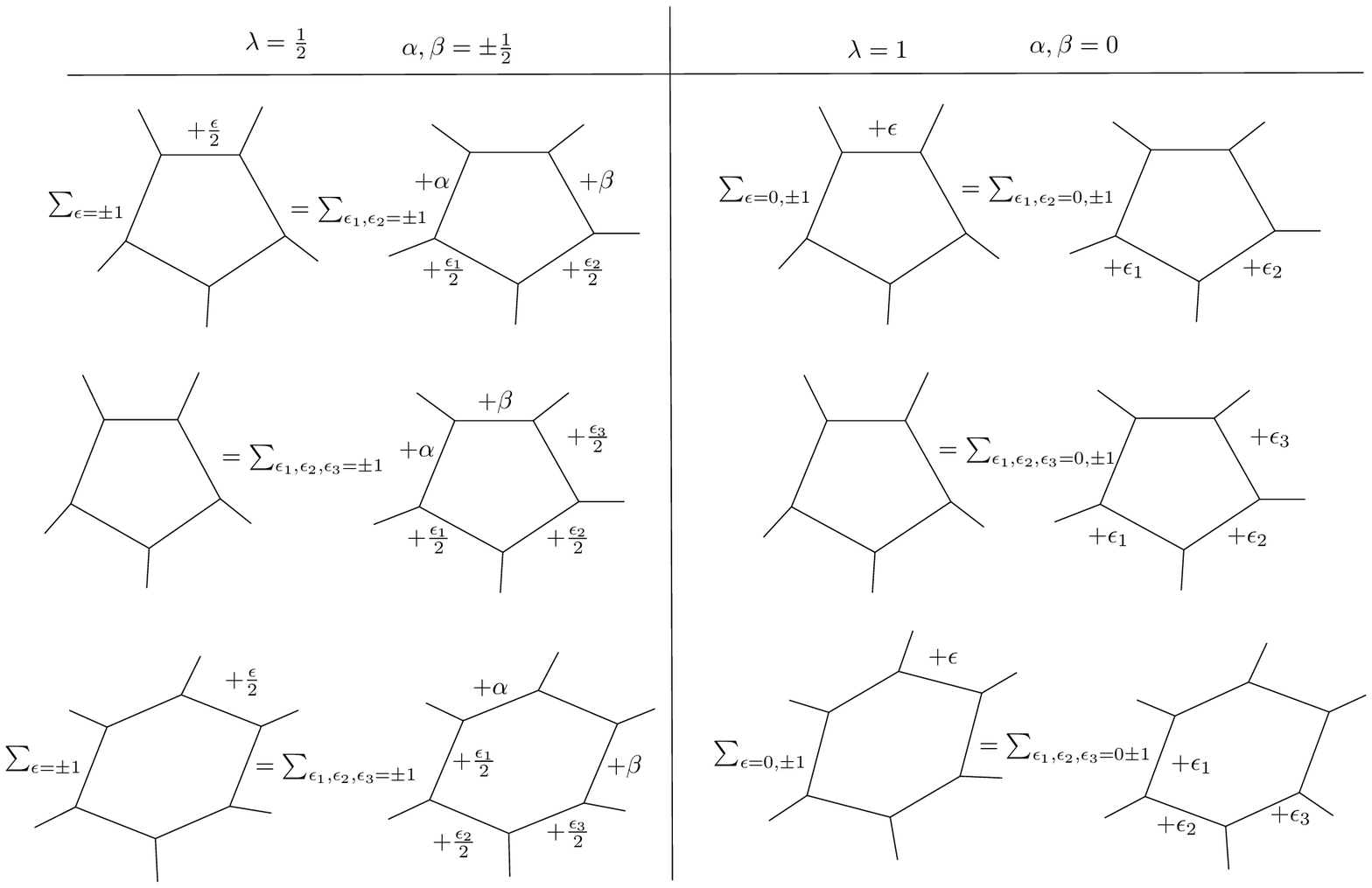}
\end{center}
\caption{ \label{fifj-summary} This array sums up the results obtained for cycles made of five and six links, with arguments changed by $\f{1}{2}$ or 1. For simplicity, the coefficients of the relations, consisting in 6j-symbols which can be explicitly evaluated, have been omitted. The recurrence relations act on any 5-cycles or 6-cycles of any $\SU(2)$ invariant graph. For $\lambda=\f 12$, one can choose the values of the free parameters $\alpha,\beta=\pm\f 12$, and all the links of the cycle are shifted. For $\lambda=1$, the choice $\alpha=\beta=0$ enables to simplify the structure of the equation by leaving some spins invariant.}
\end{figure}

These relations involve 15j-symbols whose spins are shifted by half-integers between $-\lambda$ and $\lambda$. To get explicit recurrences, the simplest choice is to take $\lambda=\f{1}{2}$. Then, the free parameters are $\alpha,\beta=\pm\f{1}{2}$, and every spin of the cycles is shifted. In each of three above formulas, exactly three triangle spins are changed, since the cycles only differ from the insertion of intertwining spins (which are also changed). The choice $\lambda=1$ is also of interest. In this case, one is free to choose $\alpha,\beta=0,\pm1$. The case $\alpha,\beta=0$ simplifies the equations since it enables to leave invariant two spins for each cycle. Geometrically, it means that only one triangle area, $j_{a'ce}$, is now changed by the move, together with two or three intertwiners. These special cases are summed up in Figure \ref{fifj-summary}.

In the above formulas, only the spins along a cycle of the spin network graphs are shifted. We emphasize that the coefficients of these relations only depend on the spins of the cycles, and on those of the links joining them. As a consequence, they apply to any invariant $\{3nj\}$-symbol containing such cycles. This gives us a generalization of the method described above for smaller symbols to symbols containing cycles of 5 and 6 links.

\medskip

Summarizing, we started from the regularized 2-4 move, and obtained recurrence relations by choosing some specific boundary spins, exactly like in 3d with the BE identity. We could also have followed another programme proposed in 3d: by acting with holonomy operators on spin network functionals. We used this method to provide a well-defined expression of the 1-4 move and at the same time get some recurrence relations. The same can be done in 4d to provide a definition of the regularized 1-5 move. In this case, the naive move diverges due to a factor $\delta(\id)^4$, which means that four delta functions are redundant. Choosing a specific spin $j_i$ among the Fourier expansion of each of these delta only contributes to the left hand side by a multiplicative factor, like for \Ref{rechol1}. Moreover it amounts to acting on the spin network functional associated to the 15j-symbol with the characters in the spin $j_i$ representations, $\prod_{i=1}^4\chi_{j_i}(G_i)$ if the elements $G_i$ are the holonomies around the four divergent faces. As we discussed in Section \ref{secRecHol}, the divergent move can be reconstructed by acting with $\prod_{i=1}^4\delta(G_i)$, leading to a precise relation between the quantization of the classical constraint $F=0$ and the 1-5 move.

We can also get this way non-trivial relations between 15j-symbols by simply fixing the four spins $j_i$. However, invariance of Ooguri's model under the 1-5 move follows from that under the 2-4 move (with orthogonalities relations) -- like in 3d, the BE identity together with an orthogonality relation can be used to prove the 1-4 invariance. Thus, the recurrence relations we would get from the 1-5 move can also be derived from the above and more general formulas.


\medskip

Although our recurrence formulas can be applied to any invariant symbol, the geometric interpretation coming from the 2-4 move only holds for 15j-symbols, due to their relation to 4d topological BF theory. Nevertheless, we may expect part of this interpretation to also hold for spin foam models aiming at describing 4d quantum gravity. Indeed, quantum states in loop quantum gravity are built with $\SU(2)$ spin network functionals, whose spins and intertwiners are eigenvalues of area  and volume operators. If a spin foam model provides the dynamics of these spin network states, we may then expect that recurrence relations for the spin foam amplitudes translate the classical constraints (i.e. symmetries) of the theory at the quantum level. Thus, the shifted spins would correspond to shifts of areas, and it is reasonable to think of describing such shifts in terms of elementary geometric moves such as the one presented here.

\section{A Recurrence Relation for the BC vertex} \label{BCrec}
In this section we consider another special $3nj$-symbol singled out by spin foam models of quantum gravity, the 10j symbol used in the BC vertex amplitude. This admits an integral representation like the isosceles 6j, and we can derive a recurrence relation for it using method (i) of inserting a constraint in the integral. This constraint is related to
a basic geometric property, namely the closure of the 4-simplex. We also propose an interpretation of the different terms entering the relation using the moves studied above.

\subsection{From the Gram matrix constraint to the recurrence relation}

We start with the expression of the Barrett-Crane 10j-symbol for a Euclidean 4-simplex as an
integral over five copies of the $\SU(2)$ group as given in
\cite{barrett,asympt,asymptlaurent,simone1}:
\be
\{10 j\}\,=\,
\int \prod_{a=1}^5 dg_a \, \prod_{a<b} \chi_{j_{ab}}(g_a^{-1}g_b).
\ee
It is a function of the ten representations $j_{ab}\in\N/2$ labeling the ten triangles of the
4-simplex. The characters $\chi_j(g)$ are defined as the trace of the group element in the
$j$-representation of $\SU(2)$ and depends entirely on its class angle $\theta$ (half the rotation
angle):
\be
\chi_j(g)\,=\,
\f{\sin d_j\theta}{\sin\theta}
\,=\,U_{2j}(\cos\theta).
\ee
The dimension of the $j$-representation  is given by $d_j=(2j+1)\in\N$. The $U_n$ are the Chebyshev
polynomials of the second kind and can be defined through the following generating functional:
\be
\sum_{n\in\N} U_n(x)t^n\,=\,\f{1}{1-2xt+t^2}.
\ee

The most useful expression of the 10j-symbol for both analytical and numerical computations
is in term of the ten class angles of the group elements $g_a^{-1}g_b$. Since these group elements
are obviously not independent, this results in a non-trivial measure on the ten class angles
$\theta_{ab}$. In fact, the measure is given simply by a constraint \cite{asymptlaurent}:
\be
\{10 j\}\,=\,
\f{4}{\pi^6}\int_0^\pi [d\theta]^{10}\, \delta(\det G[\theta]) \,\prod_{a<b}\sin
(2j_{ab}+1)\theta_{ab}.
\label{10jint}
\ee
The Gram matrix is a symmetric $5\times 5$ matrix defined as $G_{ab}=\cos\theta_{ab}$ with the
obvious rule $\theta_{aa}=0,\,\cos \theta_{aa}=1$. This constraint $\delta(G)$ contains all the
geometric information and allows to relate the 10j-symbol to the Regge amplitude for a
geometric Riemannian 4-simplex  in the large $j$ asymptotics \cite{asympt,asymptlaurent,simone1}.
We can rewrite the previous integral formula in term of the Chebyshev polynomials by defining the
ten variables $x_{ab}\,\equiv\, \cos\theta_{ab}$~:
\be
\{10 j\}\,=\,
\f{4}{\pi^6}\int_{-1}^{+1}[dx]^{10}\,\delta(\cG[x])\,\prod_{a<b}U_{2j_{ab}}(x_{ab}),
\ee
with the new symmetric Gram matrix $\cG_{ab}=x_{ab}$ for off-diagonal elements and $\cG_{aa}=1$ on
the diagonal.

We would like to use the recurrence relation on the Chebyshev polynomials to build a recurrence
relation for 10j-symbol:
$$
U_{2j+1}(x)=2xU_{2j}(x)-U_{2j-1}.
$$
This relation comes from the simple trigonometric formula:
\be
\sin(2j+2)\theta\,+\,\sin(2j)\theta\,=\, 2\cos\theta\,\sin(2j+1)\theta.
\label{trigo}
\ee
The idea is to apply this formula to the integral expression for the 10j-symbol \Ref{10jint}
and to combine the $\cos\theta$ factors so to form the determinant of the Gram matrix $\det(G)$.
Indeed the determinant reads as a sum over all permutations between five elements:
$$
\det(G)\,=\,\sum_{\sigma}\eps(\sigma)\prod_{a=1}^5\cos\theta_{a\sigma(a)},
$$
where $\eps(\sigma)$ is the signature of the permutation. We write the following recurrence
relation:
\be \label{tenj-recurrence}
0=\int [d\theta]^{10}\, \delta(\det G[\theta]) \det(G)\,\prod_{a<b}\sin (2j_{ab}+1)\theta_{ab}.
\ee
For each $\cos\theta$ factor coming the determinant, we get a linear combination of an upper and
lower shift on the corresponding representation label using equation \Ref{trigo}.

The trivial permutation $\sigma=\id$ has no $\cos\theta$ factors since $\forall a, \id(a)=a$; thus
it gives back the original 10j-symbol with no shift on the representation labels. There are
no permutations fixing only four elements. The next step are permutations fixing only three
elements, say $a=1,2,3$. Since it permutes the elements $a=4,5$, we get a factor
$-\cos^2\theta_{45}$:
$$
-\cos^2\theta\,\sin(2j+1)\theta\,=\,-\f{1}{2}\sin(2j+1)\theta\,-\,\f{1}{4}\left(\sin(2j+3)\theta+\sin(2j-1)\theta\right).
$$
We get a linear combination of the original 10j-symbol together with other 10j-symbols
where the representation label $j_{45}$ is shifted by $\pm 1$. Then come the permutations fixing
only two elements, say $a=1,2$. The action on the remaining elements $a=3,4,5$ can be either of the
two cyclic permutations on three elements. These two permutations give the same factor
$+\cos\theta_{34}\cos\theta_{45}\cos\theta_{35}$. This leads to a linear combination of upper and
lower shifts by $\pm \f12$ on the representation labels $j_{34}$, $j_{45}$ and $j_{35}$. For
permutations fixing only one element, say $a=1$, we get either have  a cyclic permutations of the
four remaining elements ($6$ possibilities) or two permutations on the two pairs ($3$
possibilities). The 4-cycles will give factors of the type
$+\cos\theta_{23}\cos\theta_{34}\cos\theta_{45}\cos\theta_{25}$ which will leads to $\pm \f12$
shifts on the relevant representation labels. On the other hand, the two pairs of permutations give
factors of the type $\cos^2\theta_{23}\cos^2\theta_{45}$ which leads to $\pm 1$ shifts on the
corresponding representation labels. Finally, we have to look at permutations that do not leave any
of the five elements invariant. Either, we have a 5-cycle ($4!$ possibilities) or a combination of
a 3-cycle and a 2-cycle ($10\times 2$ possibilities). At the end of the day, we check that:
$1+10+(2\times 10)+5\times (6+3) + (4!+20) =5!$.

This procedure leads to a recurrence relation on the 10j-symbols relating the original
amplitude with spins $j_{ab}$, amplitudes with up to five spins shifted by $\pm \f12$ and/or up to
two spins shifted by $\pm 1$ (due to 2-cycles). The complete expression involves $5!=120$ terms, so
we do not write it here. On the one hand, such a recurrence relation could be used to investigate
the asymptotics and the possible topological properties of the 10j-symbol or more generally
the behavior of (graviton) correlations in the Barrett-Crane(-like) models. However, on the other
hand, we are aware that the resulting formula can not yet be used to numerically compute the
10j-symbols by recurrence since it does not yet express a single 10j-symbol in terms of
symbols with lower spins. Under the present form, the practical use of this recurrence relation for
calculation purposes would require an infinite number of initial conditions. But we do hope to find
some trick to turn it into a recurrence relation useful for actual computations.

\medskip

Notice that the recurrence relation obtained in this way is a priori a fourth order difference equation.
We will comment on this below.

\subsection{The recurrence as a closure constraint}

The recurrence relation obtained for the BC vertex is quite complicated since it results from the expansion of a $5\times 5$ determinant. So it would be nice to identify a simple quantity satisfying this equation, at least in the asymptotical regime. Obvious solutions are given by constants, since: $\sum_\sigma \eps(\sigma) = 0$. We now look for more geometric solutions.

To this aim, we emphasize that the recurrence relation has been obtained for the 10j-symbol by inserting into its integral representation the quantity: $\det (\cos\theta_{ab})$. Its vanishing is precisely the necessary condition so that the angles $\theta_{ab}$ can be the dihedral angles of a genuine, i.e. closed, 4-simplex. Furthermore, from the ten spins, interpreted as the ten areas of a 4-simplex via the formula $A_f = d_{j_f}$, one can compute the associated dihedral angles $\bar{\theta}_{ab}(j)$ which satisfy this condition. In the asymptotics \cite{asympt,asymptlaurent,simone1}, the oscillatory part of the 10j-symbol then involves the Regge action for the 4-simplex, evaluated in terms of the areas $d_{j_f}$. Thus a natural candidate to satisfy the recurrence is the Regge action for a 4-simplex. More precisely, we will show that the cosine of the Regge action, $\cos\bigl(S(j_{ab}) + \alpha\bigr)$, as a function of the triangle areas $(2j_{ab}+1)$:
\be
S = \sum_{a<b}(2j_{ab}+1)\bar{\theta}_{ab},
\ee
where $\alpha$ an arbitrary phase, satisfies the recurrence in the asymptotics. Furthermore, the key identity to this result is indeed the closure of the 4-simplex, encoded into:
\be
\det\, \bigl(\cos\bar{\theta}_{ab}\bigr) = 0.
\ee

The recurrence relation shifts several spins together by $\pm \f 12$ according to the previous discussion. We first imagine the case of a shift of only one spin, say $j_{12}$. According to the trigonometric identity \eqref{trigo}, each shift of $\f 12$ is accompanied by an equivalent term with a shift of $-\f 12$. Also following the analysis of Schulten and Gordon about the asymptotics of the 6j \cite{schulten2}, we consider that the spins are large and expand $S_R$ to first order, neglecting its second derivatives. It gives:
\begin{align}
\f 12\cos\Bigl(S_R\bigl(j_{12}+\f 12\bigr)+\alpha\Bigr) + \f 12\cos\Bigl(S_R\bigl(j_{12}-\f 12\bigr)+\alpha\Bigr) &= \f 12\sum_{\eta=\pm 1}\cos\Bigl(S_R(j_{12})+\alpha + \eta\,\bar{\theta}_{12}(j_{12})\Bigr) \\
&= \cos\Bigl(S_R(j_{12})+\alpha\Bigr)\,\cos\bar{\theta}_{12}(j_{ab})
\end{align}
To obtain this result, we have used the Schläfli identity, $\sum_{a<b}(2j_{ab}+1)\delta\bar{\theta}_{ab} = 0$, so that the variation of $S_R$ with respect to $j_{12}$ is simply $2\bar{\theta}_{12}$.

The above expression can be easily extended to the case of an arbitrary number of shifted spins. Let us denote the shifted spins $j_{ij}$ for some $i<j$. Then, we obtain the cosine of the Regge action times the product of the cosines of the angles $\bar{\theta}_{ij}$:
\begin{align}
\f 12 \sum_{\eta_{ij}=\pm 1} \cos\Bigl(S_R\bigl(j_{ij}+\eta_{ij}\f 12\bigr)+\alpha\Bigr) &= \f 12\sum_{\eta_{ij}=\pm 1}\cos\Bigl(S_R(j_{ab})+\alpha + \sum \eta_{ij}\bar{\theta}_{ij}(j_{ab})\Bigr) \\
&= \cos\Bigl(S_R(j_{ab})+\alpha\Bigr)\prod\cos\bar{\theta}_{ij}(j_{ab})
\end{align}

Remember now that the shifts in the recurrence we are interested in are not arbitrary but correspond to permutations of the links. Moreover, the sign of each term is determined by the signature of the corresponding permutation. We thus obtain:
\be
\sum_{\sigma} \eps(\sigma)\,\cos\bigl(S_R+\alpha\bigr)\,\prod_{a=1}^5 \cos\bar{\theta}_{a\sigma(a)}
= \cos\bigl(S_R+\alpha\bigr)\ \det G[\bar{\theta}] = 0.
\ee
This obviously holds because the angles $\bar{\theta}_{ab}$ are the dihedral angles determined by the areas.

This result shows that the recurrence found for the 10j-symbol is related to a simple geometric property of a single 4-simplex, namely its closure, unlike the usual recurrence relations derived earlier via the BE identity. Let us comment on this difference.
The 10j-symbol is not invariant under any Pachner move, thus it does not correspond to the quantization of a classical topological invariant theory. Furthermore, it is also not related to general relativity \cite{Alesci3}. It does however capture some of its properties under the special restrictions of considering a single 4-simplex, and restricting attention to area observables \cite{simone1,Dan,Io}.
Therefore the recurrence relation is implementing the symmetries of such toy model.

However, this nice geometric interpretation of the recurrence relation comes from focusing on the Regge part of the asymptotics of the 10j-symbol and neglecting the other non-geometric terms \cite{asympt}. The latter turn out to dominate the asymptotics: when the spins are scaled with a large parameter $\lambda$, the Regge part goes as $\lambda^{-\f 92}$, while the dominating constribution goes as $\lambda^{-2}$ and corresponds to degenerate configurations \cite{asympt,asymptlaurent}. Further intermediate non-geometric behaviours are also observed \cite{asympt, nloBC}.
The situation is thus much richer than for the 6j: there, the geometric and non-geometric saddle points scale in the same way, and the second order recurrence relation \Ref{Hlink} includes them both. Here on the other hand the different scalings mean that recurrence relations will in general be of higher order. To see this, let us consider a simple example. We take the following asymptotic behaviour,
\be\label{totti}
\{10j\}\sim\,\f{a}{\lambda^\alpha}+\f{1}{\lambda^{\alpha+\beta}}(b\cos \lambda S+ b'\sin\lambda
S)+\dots,
\ee
where $S$ is the (scale-free) Regge action for the 4-simplex (which defines the oscillation
frequency of the 10j-symbol), and $a,b,b'$ are fixed numerical parameters.
This form is motivated by the real asymptotics of the 10j symbol with the intermediate saddle points (see \cite{asympt, nloBC}) neglected.
Then one can show that \Ref{totti} is a solution (for all parameters $a,b,b'$) of a third
order linear differential equation:
\be
\lambda(\beta(\beta-1)-aS^2\lambda^2)(\pp_\lambda^3(\lambda^{\alpha+\beta}f)+S^2\pp_\lambda(\lambda^{\alpha+\beta}f))
-\beta((\beta-1)(\beta-2)-aS^2\lambda^2)(\pp_\lambda^2(\lambda^{\alpha+\beta}f)+S^2(\lambda^{\alpha+\beta}f))
=0.
\ee
Extra terms in \Ref{totti} between the dominant term $\lambda^{-\alpha}$ and the oscillating regime will
require a higher order equation.
This example partially justifies the high order (fourth) of the recurrence relation for the 10j found above.
However, we have not been able to study the asymptotics of our recurrence relation and make such a link with the saddle points of the 10j symbol more explicit.

On the other hand, one might wonder whether restricting one's attention to geometric saddle points, second order recurrence relations with simpler geometric interpretations can be selected. There is indeed a strategy that can be used to achieve this, and which has already proved fruitful in the computations of spin foam correlations \cite{simone1,graviton,graviton3d,valentin,Alesci3}. It consists in adding a boundary state to the amplitude which naturally enhances the geometric saddle points, for instance suppressing the degenerate configurations in the bulk as in \cite{simone1}. We develop this idea in the last section.

\subsection{Geometric moves for the BC vertex}

Here we propose an interpretation of the different terms entering the relation above.
The Barrett-Crane model is not topological, so that a geometric interpretation using Pachner moves, like for the 6js and 15js, is not possible. But, since the spins labelling the 10j-symbol are quantum areas, we would like to find some elementary moves which generate the area shifts appearing in the above recurrence relation. To this aim, the picture we obtained from studying the 2-4 move for 15js seems appropriate, because we interpreted there the move through the gluing of some flattened 4-simplex, or the small displacement of a point of the initial 4-simplex, which geometrically results in some area shifts.

We will show here that if we discard what happens at intertwiners in the recurrence formulas for 15j-symbols and focus on the behaviour of spins, then the different terms of the recurrence on the 10j look very much like terms which would arise if we were working on $\SU(2)$ 15j-symbols. It provides with a qualitative picture of the recurrence as the statement of invariance under a combination of the elementary move described in Section \ref{rec2-4move} in terms of gluing some flattened simplices to the initial one. However, in contrast with the 15j case, we will not give a precise algebraic translation of these geometric considerations. In particular, we do not know the amplitude which should be assigned to the flattened simplices.


We described two elementary moves in Section \ref{rec2-4move}. First, when slightly moving a point of a 4-simplex, the spins of three triangles sharing an edge are shifted by $\f{1}{2}$, and we call it move 1 (it comes from taking $\lambda=\f 12$ in the formulas of \ref{rec2-4move}). Second, if the point is moved a little further (that is by taking $\lambda=1$), it is possible to shift the spin of only one triangle, by an amount of 1. This is move 2.  Also remember that the links of the 10j graph stands for triangles of the 4-simplex, while vertices of the graph are tetrahedra.

Let us begin with the simplest kind of shifts, when only one spin is changed by $0,\pm 1$. This happens as soon as there is a permutation of two elements in \eqref{tenj-recurrence}. Notice that no other triangle sharing an edge with the shifted triangle has its area changed. We can thus consider that such terms are obtained from the move 2. To make it precise, we use the situation described in section \ref{fifj-recurrences}, with the same notations, and in particular the figures \ref{move4-2} and \ref{move4-2graph}. We consider that the initial 4-simplex is $(abcde)$, and we proceed to a small displacement of the point $a$ to $a'$ -- for 15j-symbols, the spins labelling $(aa'c)$ and $(aa'e)$ are taken to be zero, and that of $(aa'd)$ is 1. Then, we have shown that it induces some shifts on the area of the triangle $(abd)$ (again we do not look at the algebraic level and forget the shifts on intertwiners). Clearly, this elementary move can be independently performed on triangles which do not meet at an edge.

A similar interpretation can be given to the terms consisting in a cyclic permutation of three elements. This induces shifts of $\pm\f{1}{2}$ on three links forming a triangle in the spin network graph, or equivalently on the areas of three triangles sharing an edge in the original 4-simplex. This is precisely the move 1. If in the above-mentioned flattened 4-simplex $(aa'cde)$ the spin labelling $(aa'd)$ is $\f{1}{2}$ -- or say, if it is smaller than for the move 2, since we do not have an algebraic characterization of the corresponding amplitude -- then the three triangles sharing the edge $(ad)$ are modified as expected.

Consider now the terms due to a cyclic permutation of four elements in \eqref{tenj-recurrence}, thus acting on four triangles, whose corresponding links form a cycle in the spin network graph. If we think of a diagonal joining two opposite vertices of the cycle and dividing it into two triangles, we are in position to apply twice the move 1 for each cycle of three links. As far as 15j-symbols are concerned, some shifts have constant signs which can be arbitrarily chosen. This suggests that it is possible to leave invariant the spin of the chosen diagonal link when performing the move twice, while shifting the four links of the cycle. Let us precisely see how the successive two moves act on the original 4-simplex. Consider the 4-simplex of figure \ref{move4-2} and perform the move as described there. One obtains the 4-simplex $(a'bcde)$, where the spins of $(a'bd)$, $(a'cd)$ and $(a'de)$ differ from those of $(abd)$, $(acd)$ and $(ade)$ by $\f 12$. Add now a point $a''$ close to $a'$ and consider the 4-simplex $(a'a''cde)$. Putting the spins of the triangles $(a'a''c)$ and $(a'a''d)$ to zero, and that of $(a'a''e)$ to $\f{1}{2}$, the spins of $(a''be)$, $(a''ce)$ and $(a''de)$ in the new 4-simplex $(a''bcde)$ differ from those of $(a'be)$, $(a'ce)$ (which are the same as those of $(abe)$ and $(ace)$) and that of $(a'de)$ by $\f 12$. Between the triangles $(ade)$ and $(a''de)$, the area has been affected twice, in opposite ways, so that it is unchanged. Finally it is easy to check that the affected triangles $(abd)$, $(acd)$, $(ace)$ and $(abe)$ share the point $a$, so that they form a cycle of four links in the spin network graph.

As for a cyclic permutation of the five elements in \eqref{tenj-recurrence}, the strategy is to apply thrice move 1. In the spin network graph, it affects five links forming a closed loop (a 5-cycle). It is always possible to also consider two additional links so as to form three 3-cycles, such that two of them meet at a point and both are made with two of the links of interest and one of the additional links, while the third 3-cycle consists in the two additional links and the last link closing the 5-cycle. We can then independently perform the move 1 twice to shift the six links forming the two triangles meeting at a point. The third move acts on the third 3-cycle, in order to eliminate the shifts on the additional links while changing the spin of the fifth link of interest.

\subsection{``Ward" identities for BC correlations}

We consider the framework of the ``spin foam graviton propagator" calculations on a single 4-simplex \cite{graviton,simone1,Dan,Io,Alesci3}. The boundary state is defined by the boundary probability distribution of the ten representation labels $j_1,..,j_{10}$. Here, for the sake of simplicity, we make the two usual assumptions:
\begin{itemize}

\item that the boundary state is factorisable, i.e. of the type $\prod_{i=1}^{10}\psi_i(j_i)$, where the ten functions $\psi_i$ are a priori arbitrary;

\item and that the boundary state is actually fully symmetric and ``equilateral", i.e. all the ten functions are the same, $\psi_i(j)=\psi(j), \, \forall i$, where $\psi$ is a priori still arbitrary.

\end{itemize}
For our present purpose, we do not need to require that the boundary state defined by $\psi$ be physical. We are interested in the correlation functions between the area of two triangles in the Barrett-Crane 4-simplex. To this purpose, choosing two triangles $\Delta_1=(\alpha\beta)$ and $\Delta_2=(\gamma\delta)$, we define the following expectation value:
\be
\la \cO_1(j_{\Delta_1})\cO_2(j_{\Delta_2})\ra_\psi
\,\equiv\,
\f1\cN\,
\sum_{j_{ab}}\cO_1(j_{\Delta_1})\cO_2(j_{\Delta_2})
\,\prod_{a<b}\psi(j_{ab})\,
\{10j\},
\ee
where $\cO_1$ and $\cO_2$ are the two studied observables while the normalisation $\cN$ is defined through the exact same sum without the observable insertions.
To keep things simpler, we assume that the two observables are the same, $\cO_1=\cO_2=\cO$.
%

Let us now insert the recurrence relation on the 10j-symbol into such correlations. Shifts in the 10j-symbol can be re-absorbed through the summation over representation labels by a shift of the boundary states $\psi$ and observables $\cO$. We will not write the full equation induced by the recurrence relation on the correlation functions, but we will focus on the structure of each term of that equation.

Terms with shifts on the representations will involve shifted observable insertion such as $\cO(j\pm\f12)$ and ratios between the shifted boundary state and the actual boundary state $\psi(j\pm\f12)/\psi(j)$. The standard ansatz is to look at quadratic observables and to consider phased Gaussian boundary states:
\be
\cO(j)=j^2\,\Rightarrow\,\cO(j\pm\f12)=\cO(j)\pm j+\f14,
\ee
\be
\psi(j)=e^{id_j\vtheta}e^{-\alpha (j-j_0)^2/j_0}
\,\Rightarrow\,
\f{\psi(j\pm\f12)}{\psi(j)}\,=\,
e^{i\vtheta}e^{-\f{\alpha}{4j_0}}e^{\pm\alpha\left(1-\f{j}{j_0}\right)},
\ee
where $\vtheta$ and $j_0$ are the fixed parameters of the boundary states: $\vtheta$ correspond to the classical (exterior) dihedral angle while $j_0$ defines the scale factor of the 4-simplex (its size). And $\alpha$ defines the width of the Gaussian state.
This way, we derive an equation between the 2-point function, the 3-point function, up to the 5-point function, with different observable insertions corresponding to shifts in the original observable or boundary state. It's a type of Ward identity for Barrett-Crane correlations on the 4-simplex. However, it is a priori not straightforward to interpret it as a function describing the scaling of the correlation function with changes of the scale factor $j_0$.

\section{Conclusion}

In this paper, we have derived recurrence relations for 6j-symbols and 15j-symbols. For 6j-symbols, we used different methods: the standard Biedenharn-Elliott identity which correponds to the 2-3 Pachner move, and the action of holonomy operators on spin network functionals. This second method naturally generates a regularized 1-4 Pachner move and gives a precise relation to the classical constraints of the theory. Then, in 4d BF theory, recurrence relations for 15j-symbols have been derived from a regularized 2-4 Pachner move. The invariance of the Ponzano-Regge and Ooguri models under these Pachner moves is due to their underlying topological invariance, and provides the recurrence relations with a nice geometric picture: the shifts on the spins, which are eigenvalues of lengths and fluxes operators in the canonical theories, are generated by adding a flattened simplex to an initial one, resulting in a small displacement of a point of this simplex. We hope that such a picture will survive in non-topological theories and help to describe the action of the Hamiltonian constraints and to analyze the behavior of spin foam amplitudes under coarse-graining.

Interestingly, the recurrences we found can be applied to any $\SU(2)$ $\{3nj\}$-symbol, depending on the form of their cycles. Indeed, we found that the relations naturally shift the spins along one cycle of the graph. Moreover, the form and the coefficients of the relations only depend on the number of links of the cycle. We found relations for cycles made of up to six links. Besides its theoritical interest, we think that this feature could be useful to speed up numerical computations.

For the special isosceles 6j-symbols, a new recurrence relation has been obtained using an integral representation. We extracted its geometric content by also deriving it from the BE identity. Then, we applied the same method to the 10j-symbol which is the 4-simplex amplitude of the Barrett-Crane model. While the recurrence formulas for the 6j and 15j-symbols were difference equations of order 2, we found there a difference equation of order 4 for the 10j-symbol. This aspect is certainly related to the following fact. The evaluation of the 10j-symbol in the asymptotics, like that of the 6j-symbol, using the saddle point approximation \cite{asympt,asymptlaurent} exhibits different kinds of saddle points: a geometric one together with non-geometric (degenerate) contributions. Although these different terms have the same scaling for the 6j, this is not the case anymore for the 10j, resulting in an increasing of the order of the difference equation. To get rid of the non-geometric saddle points, we then looked for recurrence equations for the vertex amplitude alone put together with a geometric boundary state and derived some Ward-like identities for insertions of geometric observables.

We have also shown that the cosine of the Regge action solves the recurrence equation in the asymptotics. But we could not yet show that the other terms corresponding to degenerate saddle points also solve our recurrence formula, since we do not know an explicit formula for them. Looking at the Regge part of the asymptotics, the recurrence relation takes the form of a constraint imposing the closure of the 4-simplex. While the BC model is not topological, we have also proposed a qualitative picture to describe the various terms entering the recurrence, with the help of the elementary geometric moves which have emerged from the analysis of the 15j-symbol.

Although the BC model  probably does not provide a correct quantization of gravity, it is certainly a useful toy model to try to extend the results about spin foam models beyond the topological case. To complete the analysis of the 10j, it would be of great interest to derive the recurrence we found from a coarse-graining process, similarly to what we have done with the 15j-symbol. We also hope to apply the same methods to the more complicated Engle-Pereira-Rovelli \& Freidel-Krasnov models \cite{epr}, which have a better semi-classical behavior. However, the intertwiners are then not frozen anymore and have to be taken into account in the recurrence relation and in the coarse-graining process. A first step in that direction could be to rewrite and reinterpret the 15j relations in term of the coherent state representation \cite{coherent, coherent2}.


\appendix

\section{The usual recurrence for the \sj-symbol}
6j-symbols admit a recurrence relation of the following simple form:
\begin{multline}
j_1 E(j_1+1)\,\begin{Bmatrix} l_1 &l_2 &l_3 \\ j_1+1 &j_2 &j_3 \end{Bmatrix} + \bigl(2j_1+1\bigr)\Bigl\{2\bigl[j_2(j_2+1)l_2(l_2+1)+j_3(j_3+1)l_3(l_3+1)-j_1(j_1+1)l_1(l_1+1)\bigr] \\- \bigl[j_2(j_2+1)+l_3(l_3+1)-j_1(j_1+1)\bigr]\bigl[l_2(l_2+1)+j_3(j_3+1)-j_1(j_1+1)\bigr]\Bigr\}\,\begin{Bmatrix} l_1 &l_2 &l_3 \\ j_1 &j_2 &j_3 \end{Bmatrix} + \bigl(j_1+1\bigr)E(j_1)\,\begin{Bmatrix} l_1 &l_2 &l_3 \\ j_1-1 &j_2 &j_3 \end{Bmatrix} = 0,
\end{multline}
where:
\be
E(j_1) = \bigr[(j_1+j_2+l_3+1)(j_1-j_2+l_3)(j_1+j_2-l_3)(-j_1+j_2+l_3+1)(j_1+j_2+j_3+1)(j_1-l_2+j_3)(j_1+l_2-j_3)(-j_1+l_2+j_3+1)\bigr]^{\f{1}{2}}.
\ee
It directly comes from taking one boundary spin to 1 in the Biedenharn-Elliott identity, and evaluating explicitly the 6j-symbols which involve the spin 1. In the large spin regime, it can be turned into a second order differential equation, whose analysis leads to the known asymptotics of the 6j-symbol \cite{schulten2}.

\section{Some remarks on the Chebyshev polynomials}

\be
U_{2j}(x)\,=\,\sum_{k=0}^{\lfloor j\rfloor}(-1)^k
\left(\begin{array}{c}2j-k\\k\end{array}\right) (2x)^{2(j-k)}.
\ee

\be
U_{2j}(x)\,=\,2^{2j}\prod_{k=1}^{2j}\left(x-\cos\f{k\pi}{2j+1}\right).
\ee

\section{Generating Functional for the BC vertex}

We can also study the generating functional for the Barrett-Crane vertex. Indeed, using its
expression in term of the Chebyshev polynomials $U_n$, we introduce the following function of $10$
arguments $t_{ab}$ for $1\le a<b\le 5$:
\be
BC(t_{ab})\,\equiv\,
\sum_{j_i\in\N/2} \prod_{a<b} t_{ab}^{2j_{ab}}\,\{10j\}
\,=\,
\f{4}{\pi^6}\int_{-1}^{+1}[dx]^{10}\,\delta(\cG[x])\,
\prod_{a<b}\f{1}{1-2x_{ab}t_{ab}+t_{ab}^2}.
\ee
This is the integral of a rational fraction with a constraint. It looks like the evaluation of a
Feynman diagram and should be computable with similar methods (contour integrals). Or else finding
a differential equation satisfied by $BC[t]$ should lead to recurrence relations on the
10j-symbols. Moreover, this generating functional could be directly relevant to the
computation of correlation functions following the framework on the spinfoam graviton
\cite{simone1, graviton}.



\end{document}